\documentclass[%
 aip,
 amsmath,amssymb,
 reprint,%
]{revtex4-1}

\usepackage{graphicx}
\usepackage{dcolumn}
\usepackage{bm}

\usepackage[utf8]{inputenc}
\usepackage[T1]{fontenc}
\usepackage{mathptmx}

\begin{document}

\preprint{AIP/123-QED}

\title[Estimating entropy rate from censored symbolic time series]{Estimating entropy rate from censored symbolic time series: a test for time-irreversibility}

\author{R. Salgado-Garc{\'\i}a}
\email{raulsg@uaem.mx}
\affiliation{Centro de Investigaci\'on en Ciencias-IICBA, Physics Department, Universidad Aut\'onoma del Estado de Morelos. Avenida Universidad 1001, colonia Chamilpa, CP 62209, Cuernavaca Morelos, Mexico. }

\author{Cesar Maldonado}
\affiliation{IPICYT/Divisi\'{o}n de Control y Sistemas Din\'{a}micos. Camino a la Presa San Jos\'{e} 2055, Lomas 4a. secci\'{o}n, C.P. 78216, San Luis Potos\'{i}, S.L.P. Mexico. }

\date{\today}

\begin{abstract}
In this work we introduce a method for estimating entropy rate and entropy production rate from finite symbolic time series. From the point of view of statistics, estimating entropy from a finite series can  be interpreted as a problem of estimating parameters of a distribution with a censored or truncated sample. We use this point of view to give estimations of entropy rate and entropy production rate assuming that they are parameters of a (limit) distribution. The last statement is actually a consequence of the fact that the distribution of estimations obtained from recurrence-time statistics satisfy the central limit theorem. We test our method using time series coming from Markov chain models, discrete-time chaotic maps and real a DNA sequence from human genome. 
\end{abstract}

\maketitle

\begin{quotation}

Entropy rate as well as entropy production rate are fundamental properties of stochastic processes and  deterministic dynamical systems. For instance, in dynamical systems the entropy rate is closely related to the largest Lyapunov exponent, stating that the positivity of entropy rate is a signature of the presence of chaos. Similarly, the entropy production rate is a measure of the degree of irreversibility of a given system. Thus, in some sense, a non zero entropy production rate states how much, a system, is far from equilibrium. However, estimating either, entropy rate or entropy production rate is not a trivial task. One of the main limitations to give precise estimations of these quantities is the fact that observed data (time series) are always finite, but the entropy rate and entropy production rate are asymptotic quantities defined as a limit for which it is necessary to have infinitely long time-series. We use the recurrence-time statistics combined with theory of censored samples from statistics to propose sampling schemes and define censored estimators for the entropy rate and the entropy production rate, taking advantage of the finiteness of the observed data.

\end{quotation}

\section{\label{sec:intro}Introduction}


Entropy rate and entropy production rate are two quantities playing a central role in equilibrium and nonequilibrium statistical mechanics. On the one hand, entropy rate (also called Kolmogorov-Sinai entropy) is closely related to the thermodynamical entropy~\cite{Cornfeld1989,latora1999kolmogorov} which is a fundamental quantity in the context of equilibrium statistical mechanics. On the other hand, entropy production has a prominent role in the development of nonequilibrium statistical mechanics~\cite{gallavotti2014nonequilibrium,RM,Mae}. Both, entropy rate and entropy production rate, have a rigorous definition in dynamical systems and stochastic processes (see Ref.~\onlinecite{Jiang} for complete details). The entropy production rate quantifies, in some way, the degree of time-irreversibility of a given system from a microscopic point of view, which in turn tells us how much such a system is far from the thermodynamic equilibrium~\cite{RM,gaspard2004time,Mae}. Moreover, time-irreversibility of certain dynamical processes in nature might be an important feature because it would imply the influence of nonlinear dynamics or non-Gaussian noise on the dynamics of the system ~\cite{daw2000symbolic}. All these features of time-irreversibility has encouraged the study of this property in several systems. For instance, in Ref.~\onlinecite{Provata2014} it has been found that real DNA sequences would be spatially irreversible, a property that has been explored aimed to understand the intriguing statistical features of the actual structure of the genome. The fact that DNA might be spatially irreversible has been used to propose a mechanism of noise-induced rectification of particle motion~\cite{rsg2019noise} that would be important in the study of biological processes involving the DNA transport.  Testing the irreversibility of time series has also been the subject of intense research. For example, in Ref.~\onlinecite{daw2000symbolic} it has been proposed a symbolic dynamics approach to determine whether the time series are time-irreversible or not. Another important study has been reported in Ref.~\onlinecite{lacasa2012time}, where the authors introduced a method to determining time-irreversibility of time series by using a visibility graph approach. That  approach has also been used to understanding the time-reversibility of non-stationary processes~\cite{lacasa2015time}. The possibility of determining this temporal asymmetry has also lead to try to understand the dynamics of several processes beyond physical systems. In Ref.~\onlinecite{flanagan2016irreversibility} it has been explored the time-irreversibility of financial time-series as a feature that could be used for ranking companies for optimal portfolio designs.  In Ref.~\onlinecite{costa2005broken} it has  been studied the time-irreversibility of human heartbeat time-series, and relating this property to aging and disease of individuals. Moreover, time-irreversibility  has also been used to understand several properties of classical music~\cite{Gustavo}.

In the literature one can find many estimators of the entropy rate in symbol sequences produced by natural phenomena as well as in dynamical systems, random sequences or even in natural languages taken from written texts. Perhaps, the most used method for entropy estimation is the empirical approach, in which one estimates the probability of the symbols using their empirical frequency in the sample and then, this is used to estimate the entropy rate directly from its definition~\cite{Bonachela,SchGra}. One can find a lot of works in this direction trying to find better, unbiased and well-balanced estimators (see Ref.~\onlinecite{Schurmann} and references therein). One can go further by asking for the consistency and the fluctuation properties of these estimators. For instance in Refs.~\onlinecite{AK} and~\onlinecite{ChM} there are explicit and rigorous fluctuation bounds under some mild additional assumptions, for these so-called ``Plug-In" estimators. On the other hand, but in the same empirical approach, there are also estimators for the relative empirical entropy as a quantification of the entropy production~\cite{PorpoAl,RolP}. 

From another point of view, the problem of estimating the entropy rate of stationary processes has also been studied using the recurrence properties of the source. This is, another major technique used in the context of stationary ergodic processes on the space of infinite sequences, in areas such as information theory, probability theory and in the ergodic theory of dynamical systems (we refer the interested reader to Ref.~\onlinecite{shields} and the references therein ). The basis of this approach is the Wyner-Ziv-Ornstein-Weiss theorem which establishes an almost sure asymptotic convergence of the logarithm of the recurrence time of a finite sample (scaled by its length), to the entropy rate~\cite{shields}. This result uses the Shannon-McMillan-Breiman theorem, which in turn, can be thought as an ergodic theorem for the entropy~\cite{shields}. Under this approach it is possible to define estimators using quantities such as return time,  hitting time,  waiting time among others~\cite{AG}. Here we will use the term ``recurrence time'' as a comprehensive term for those mentioned before. Moreover, it is possible to obtain very precise results on the consistency and estimation of the fluctuations of these estimators by applying the available results on the distribution of these quantities~\cite{Aba,AV,Kon}.

In the setting of Gibbs measures in the thermodynamic formalism, one can also find consistent estimators defined from the return, hitting, and waiting times for entropy rate and one also has precise statements on their fluctuations, such as the central limit theorem~\cite{ChU}, large deviation bounds and fluctuation bounds~\cite{ChU,ChM}. Similarly occurs within the study of the estimation of the entropy production rate.  In the context of Markov chains applied to the quantification of the irreversibility or time-reversal asymmetry see Refs.~\onlinecite{gaspard2004time,CofreM}, in Gibbssian sources see Ref.~\onlinecite{ChR} as well as for their fluctuation properties in Ref.~\onlinecite{ChR,cesar2015fluctuations}.

Nonetheless, for real systems, determining the value of the entropy rate and the entropy production rate is not a trivial task. This is because these quantities are obtained as limit values of the logarithm of recurrence times, as the sample length goes to infinity. This is a fundamental limitation, since observations are always finite. So, instead of having the true value of the entropy rate or the entropy production rate, one always obtains a finite-time approximation. This makes us believe that there is a need to define estimators  for finite samples, using the point of view of the recurrence times. 

The article is organized as follows. In Section~\ref{sec:entropy} we give a summary of the asymptotic properties of the estimators based on the recurrence-time statistics. We also describe the method used for estimating parameters of the normal distribution from a given censored sample. In Section~\ref{sec:estimating_entropy} we propose our sampling schemes for estimating the entropy rate and the reversed entropy rate using the recurrence-time statistics. There, we also describe the method that will be used for implementing the estimations in real data. In Section~\ref{sec:tests} we test the methodology established in Section~\ref{sec:estimating_entropy} for estimating the entropy rate and the reversed entropy rate in an irreversible three-state Markov chain. We compare our estimations with the exact values that can be actually computed. In Section~\ref{sec:examples} we implement the proposed estimating method in deterministic chaotic systems, a $n$-step Markov chain and a real DNA sequence. Finally in Section~\ref{sec:conclusions} we give the main conclusions of our work.

\section{\label{sec:entropy}Entropy rate and entropy production rate}


\subsection{Recurrence time statistics}

Consider a finite set $A$ which we will refer to as \emph{alphabet}. Let $\mathbf{X} := \{ X_{n} \, : \, n\in \mathbb{N}\}$ a discrete-valued stationary ergodic process generated by the law $\mathbb{P}$, whose realizations are infinite sequences of symbols taken from $A$, that is, the set of all posible realizations is a subset of $A^{\mathbb{N}}$. Here we denote by $\mathbf{x} = x_{1} x_{2} x_{3}\ldots$ an infinite realization of the process $\mathbf{X}$. Let  $\ell$ be a positive integer, we denote by $x_{1}^{\ell}$ the string of the first $\ell$ symbols of the realization $\mathbf{x}$. A finite string $\mathbf{a} := a_1 a_2a_3\ldots a_\ell$ comprised of $\ell$ symbols will be called either $\ell$-word or $\ell$-block, we may use one or the other without making any distinction. We will say that the $\ell$-word $\mathbf{a}$ ``occurs'' at the $k$th site of the trajectory $\mathbf{x}$, if $x_k^{k+\ell-1} = \mathbf{a}$. An alternative notation for indicating the $\ell$-block at the $k$th site of  $\mathbf{x}$ will be: $\mathbf{x}(k,k+\ell-1)$.

Next, we introduce the return time, the waiting time and the hitting time. Let us consider a finite string $a_{1}^{\ell}$ made out of symbols of the alphabet $A$. Given two independent realizations $\mathbf{x}$ and $\mathbf{y}$,  let $x_{1}^{\ell}$ and $y_{1}^{\ell}$ be their first $\ell$ symbols, then the return, the waiting and the hitting time are defined as follows,
\begin{equation}
\rho_{\ell} := \rho_{\ell}(\mathbf{x}) := \inf\{ k> 1\ :\ x_{k}^{k+\ell-1} = x_{1}^{\ell}\},
\end{equation}
\begin{equation}
\omega_{\ell}:= \omega_{\ell}(\mathbf{x}, \mathbf{y}) := \inf \{ k\geq 1\ :\ y_{k}^{k+\ell-1} = x_{1}^{\ell} \},
\end{equation}
\begin{equation}
\tau_{\ell} := \tau_{\ell}(a_{1}^{\ell},\mathbf{x}) := \inf\{ k\geq 1 \ :\ x_{k}^{k+\ell-1}=a_{1}^{\ell} \},
\end{equation}
 respectively.

Wyner and Ziv (see for instance Ref. \onlinecite{WZ}) proved that for an stationary ergodic process, the quantity $\frac{1}{\ell} \log \rho_{\ell}$ converges to the entropy rate in probability, and that for stationary ergodic Markov chains,  $\frac{1}{\ell} \log \omega_{\ell}$ also converges to the entropy rate $h$, in probability. That is, these quantities grow exponentially fast with $\ell$ and their limit rate is equal to the entropy rate in probability. Later, Ornstein and Weiss~\cite{OW} showed that for stationary ergodic processes
\begin{equation}
\lim_{\ell\to\infty}\frac{1}{\ell} \log\rho_{\ell}  = h \quad \quad \mathbb{P}-\mbox{a.s.}
\end{equation}
For the waiting time, it was proved by Shields~\cite{shields} that for stationary ergodic Markov chains one has,
\begin{equation}
\lim_{\ell\to\infty}\frac{1}{\ell}\log \omega_{\ell} = h \quad \quad \mathbb{P}\times \mathbb{P}-\mbox{a.s.}
\end{equation}
These theorems are based on the Shannon-McMillan-Breiman theorem, which claims that $-\frac{1}{\ell} \log \mathbb{P}([x_{1}^{\ell}])$ converges almost surely to the entropy rate $h$, where $[x_{1}^{\ell}]$ stands for the cylinder set $[x_{1}^{\ell}]:=\{\mathbf{z}\in A^{\mathbb{N}} \ :\ z_{1}^{\ell} = x_{1}^{\ell}\}$. Furthermore, in Ref.~\onlinecite{Kon}, Kontoyiannis has obtained strong approximations for the recurrence and waiting times of the probability of a finite vector which in turn, have let him to obtain an almost sure convergence for the waiting time in $\psi$-mixing processes, extending previous results for Markov chains. He has also obtained an almost sure invariance principle for $\log \rho_{\ell}$ and $\log \omega_{\ell}$. This implies that these quantities satisfy a central limit theorem and a law of iterated logarithm.~\cite{Kon}

In the same spirit, the works of Abadi and collaborators~\cite{Aba,AV,AG} provide very precise results for the approximation of the distribution of the hitting times (properly rescaled) to an exponential distribution, under mild mixing conditions for the process. They also give sharp bounds for the error term for this exponential distribution approximation. This enables to obtain bounds for the fluctuations of the entropy estimators using hitting times.~\cite{ChU,ChM,cesar2015fluctuations}

\subsection{Asymptotic behavior of the estimators}

We are interested in estimating the entropy and the entropy production rates, moreover, we need to assure that their estimators have good properties of convergence and fluctuations, since this will enable us to use our method.

Here, we are interested in estimators defined by recurrence times, for which one can find very precise asymptotic results regarding their fluctuations. It is known~\cite{OW} that 
\begin{equation}
\lim_{\ell\to\infty}\frac{1}{\ell}\log \rho_{\ell} = h,
\end{equation}
almost surely in ergodic process, thus one can use the return time as an estimator of the entropy rate. Furthermore, under the Gibbssian assumption, it has been proved that the random variable $(\log \rho_{\ell} - \ell h)/\sqrt{\ell}$ converges in law to a normal distribution, when $\ell$ tends to infinite.~\cite{CGS}

The waiting and hitting times are also used as estimators. For instance, it has beed proved that,
\begin{equation}
\lim_{\ell\to\infty} \frac{1}{\ell}\log \omega_{\ell}(\mathbf{x}, \mathbf{y}) = h,
\end{equation}
for $\mathbb{P}\times\mathbb{P}$ almost every pair $(\mathbf{x}, \mathbf{y})$, where the distribution $\mathbb{P}$ is a Gibbs measure.~\cite{ChU}
This is obtained from an approximation of the $\frac{1}{\ell}\log \omega_{\ell}$ to the $-\frac{1}{\ell}\log\mathbb{P}([x_{1}^{\ell}])$ which, by the Shannon-McMillan-Breiman theorem, goes almost surely to the entropy rate. Also, they proved the same log-normal fluctuations for the waiting times, i.e.,
\begin{equation}
\lim_{\ell\to\infty} \mathbb{P}\times\mathbb{P}\Big\{\frac{\log \omega_{\ell} - \ell h}{\sigma \sqrt{\ell}}< t \Big\} = \mathcal{N}(0,1)(-\infty,t],
\end{equation}
where in this case, $\sigma^{2}= \lim_{\ell\to\infty}\frac{1}{\ell}\int (\log \omega_{\ell} - h)^{2}\textup{d}(\mathbb{P}\times\mathbb{P})$.

So, in the context of Gibbs measures, the asymptotic normality it is fulfilled for both, the return times and the waiting times. This also holds for exponential $\phi$-mixing processes. Moreover it is satisfied a large deviations principle for both quantities as well~\cite{ChU} (with some additional restrictions in the case of the return-time). For the case of the hitting times, one has to overcome the bad statistics produced by very short returns for which the approximation changes (see Ref.~\onlinecite{AL}).

In the same context, one can find fluctuation bounds for both, the plug-in estimators and for the waiting and the hitting time estimators.~\cite{ChM} One of the main tools used is the concentration inequalities that are valid for very general mixing processes. Using the concentration phenomenon, one can obtain non-asymptotic results. That is, upper bounds for the fluctuations of the entropy estimator which are valid for every $n$, where $n$ denotes the length of the sample.

Next, for the estimation of the entropy production rate, in Ref.~\onlinecite{ChR}, two estimators of the entropy production were introduced. The entropy production was defined as a trajectory-valued function quantifying the degree of irreversibility of the process producing the samples, in the following way: let $\mathbb{P}$ be the law of the process and let us denote by $\mathbb{P}_{\mathrm{r}}$ the law of the time-reversed process, then the entropy production rate is the relative entropy rate of the process with respect to the time-reversed one, 
\begin{equation}
e_{\mathrm{p}} = h(\mathbb{P}| \mathbb{P}_{\mathrm{r}}) := \lim_{\ell\to\infty}\frac{H_{\ell}(\mathbb{P}|\mathbb{P}_{\mathrm{r}})}{\ell},
\end{equation}
where
\begin{equation}
H_{\ell}(\mathbb{P}|\mathbb{P}_{\mathrm{r}}) := \sum_{x_{1}^{\ell}\in A^{\ell}} \mathbb{P}([x_{1}^{\ell}])\log\frac{\mathbb{P}([x_{1}^{\ell}])}{\mathbb{P}([x_{\ell}^{1}])}.
\end{equation}
Here $x_{\ell}^{1}$ stand for the word $x_{1}^{\ell}$ reversed in order. 
The estimators defined in Ref.~\onlinecite{ChR} using the hitting and waiting times are given as follows:
\begin{equation}
\mathcal{S}_{\ell}^{\tau}(\mathbf{x}) := \log \frac{\tau_{x_{\ell}^{1}}(\mathbf{x})}{\tau_{x_{1}^{\ell}}(\mathbf{x})},
\end{equation}
where $\tau_{x_{1}^{\ell}}(\mathbf{x}) := \inf\{k\geq 1 : x_{k}^{k+\ell} = x_{1}^{\ell}\}$. Notice that the estimator actually quantifies the logarithm of the first time the word $x_{1}^{\ell}$ appears in the reversed sequence divided by the first return time of the first $\ell$ symbols in $\mathbf{x}$. For the case of the estimator using the waiting time, one has in an analogous way that:
\begin{equation}
\mathcal{S}_{\ell}^{\omega}(\mathbf{x},\mathbf{y}) := \log \frac{\omega_{\ell}^{r}(\mathbf{x},\mathbf{y})}{\omega_{\ell}(\mathbf{x},\mathbf{y})},
\end{equation}
where $\omega_{\ell}(\mathbf{x},\mathbf{y}) := \tau_{x_{1}^{\ell}}(\mathbf{y})$ and $\omega_{\ell}^{r}(\mathbf{x},\mathbf{y}) : = \tau_{x_{\ell}^{1}}(\mathbf{y})$. In the context of Gibbs measures or exponential $\psi$-mixing,~\cite{ChR} it has been studied the fluctuation properties of such estimators for which its consistency has also been proved, that is, $\mathbb{P}\times\mathbb{P}$-almost surely we have that,
\begin{equation}
\lim_{\ell\to\infty} \frac{\mathcal{S}_{\ell}^{\omega}}{\ell} = e_{\mathrm{p}},
\end{equation}
as well as, $\mathbb{P}$-almost surely
\begin{equation}
\lim_{\ell\to\infty} \frac{\mathcal{S}_{\ell}^{\tau}}{\ell} = e_{\mathrm{p}}.
\end{equation}
The asymptotic normality also holds, in that case, the asymptotic variance of the estimator coincides with that of the entropy production. In the same reference the authors also obtain a large deviation principle for the waiting time estimator. Later in Ref.~\onlinecite{cesar2015fluctuations} the fluctuation bounds were obtained for the same estimators introduced in Ref.~\onlinecite{ChR} under the same setting. This result is interesting from the practical point of view since it provides bounds that are valid for finite time and not only in the asymptotic sense.

Here, we will use the approach defined in Ref.~\onlinecite{gaspard2004time} for the estimation of the entropy production rate, since we want to compare it with the exact results one is able to obtain for Markov chains. In Ref.~\onlinecite{gaspard2004time} it is shown that the entropy production rate can be obtained as the difference between the entropy rate and the \emph{reversed} entropy rate for Markov processes. For more general systems, the entropy production is defined in some analogous way.~\cite{Mae} The reversed entropy rate is defined as the rate of entropy of the reversed process in time, i.e., as if we were estimating the entropy rate of the process evolving backwards in time. From the practical point of view, in a time series, the entropy production rate may be estimated as the difference between  the entropy rate and the entropy rate estimated from the reversed time series. To implement the latter methodology using the recurrence time statistics, in Section~\ref{sec:estimating_entropy} we will  define the reversed recurrence times which will allow us to give estimations of the reversed entropy rate and eventually, the corresponding estimations of the entropy production rate as a measure of time-irreversibility of the process.  It is important to mention that our methodology can still be applied further than Markov chains, nevertheless, in those cases, one expects to obtain results displaying the irreversibility as a consequence of the positivity of the entropy production, and not the exact results.

\subsection{\label{ssec:estimation_normal}Parameter estimation of a  normal distribution from censored data}


Let us denote by $\Theta_\ell$ the random variable whose realizations are estimations of the $\ell$-block entropy rate obtained by the recurrence-time statistics. To be precise, $\Theta_\ell $ can be defined as
\begin{equation}
\label{eq:def:Hell}
\Theta_\ell = \frac{1}{\ell} \log(T_\ell),
\end{equation}
where $T_\ell$ can be the return, hitting, or waiting time random variable. As pointed out above, $\Theta_\ell$ satisfy the central limit theorem regardless the choice of the recurrence time statistics. This fact enables us to assume that $\Theta_\ell$ has a normal distribution, with mean $h_\ell$ and variance $\sigma^2_\ell$.  As mentioned before, one of the problems arising in implementing this estimator for real time series is that the return time $T_\ell$ is censored from above by a prescribed finite value $T_c$ . From eq.~(\ref{eq:def:Hell}), it is clear  that the random variable $\Theta_{\ell}$ becomes censored from above by a finite value $h_c := \log(T_c)/\ell$ which will be referred to as \emph{censoring entropy}. Taking into account this observation, we can state our problem as follows: given a sample set $\{h_i: 1\leq i \leq m\}$ of independent realizations of $\Theta_\ell$, we wish to estimate $h_\ell$ and $\sigma_\ell$ knowing that such a sample is censored from above by $h_c$. 

It is important to remark that, since the realizations of $\Theta_\ell$ are censored from above by $h_c$, then any sample set $\mathcal{H}:=\{h_i: 1\leq i \leq m\}$ of (independent) realizations of $\Theta_\ell$ will contain \emph{numerically undefined} realizations; i.e., $h_i$ such that $h_i >h_c$. We well refer to these numerically undefined values as \emph{censored realizations} or \emph{censored samples}. Those sample with a well-defined numerical value will be called  \emph{uncensored} samples o realizations. We will see below that censored sample data will be used for the estimation of $h_\ell$ and $\sigma_\ell$. 

Let  $m:=|\mathcal{H}|$ be the size of the sample and let us assume that the total number of uncensored realizations  in the sample set $\mathcal{H}$ is exactly $k$, with $k<m$. Then, the total number of censored realizations in $\mathcal{H}$ is $m-k$. Since the realizations are assumed to be independent (a usual hypothesis in statistics), we have that  $k$ can be seen as a realization of a random variable with binomial distribution. Thus, the fraction $\hat p := k/m$ of uncensored samples with respect to the total number of realizations in $\mathcal{H}$ is an estimation of the parameter $p$ of the above-mentioned binomial distribution. As we said above, $\Theta_\ell$ has normal distribution, implying that the parameter $p$ is given by,
\begin{equation}
p =  \Phi\left(\frac{h_c - h_\ell}{\sigma_\ell}\right),
\end{equation} 
where $\Phi$ is the distribution function of a standard normal random variable, i.e.,
\begin{equation}
\Phi (x) := \frac{1}{\sqrt{2\pi}}\int_{-\infty}^x  e^{-y^2/2}dy.
\end{equation}

In Appendix~\ref{ape:censored}, following calculations from Ref.~\onlinecite{cohen1991truncated}, we show that the parameters $h_{\ell}$ and $\sigma_\ell^2$ can be estimated by using the censored sample as follows:
\begin{eqnarray}
\label{eq:hat_h}
\hat h &=& \bar{h} +\hat \zeta (h_c-\bar{h}),
\\
\label{eq:hat_sigma2}
\hat \sigma^2 &=& s^2 + \hat \zeta (h_c-\bar{h})^2,
\end{eqnarray}
where  $\bar{h}$ is the sample mean of the uncensored samples and $s^2$ the corresponding sample variance, i.e.,
\begin{eqnarray}
\bar{h} &:=& \frac{1}{k}\sum_{i=1}^{k} h_i,
\label{eq:est-h}
\\
s^2 &:=& \frac{1}{k} \sum_{i=1}^{k} (h_i-\bar{h})^2.
\label{eq:est-sigma}
\end{eqnarray} 
Additionally $\hat \zeta$ is defined as:
\begin{equation}
\hat \zeta := \frac{\phi(\hat \xi)}{\hat{p}\, \hat \xi + \phi(\hat \xi)}
\end{equation}
where $\hat \xi$ is obtained by means of the normal distribution function as
\begin{equation}
\hat \xi := \Phi^{-1}(\hat p).
\end{equation}

\section{\label{sec:estimating_entropy}Sampling schemes for estimating entropy rate from recurrence-time statistics}


As we said above, we are interested in estimating the entropy rate and the entropy production rate from an observed trajectory. The trajectory, in this context, stands for a finite-length symbolic sequence $\mathbf{x} = x_1x_2x_3\dots x_n$ which is assumed to be generated by some process with an unknown law $\mathbb{P}$. As we saw in section~\ref{sec:entropy}, we have to assume that the process complies with the appropriate mixing properties, such as exponential $\phi$-mixing or Gibbs, in order for the central limit theorem to be valid. The next step is to obtain samples of the recurrence time statistics, i.e., we need to establish a protocol for extracting samples of return, waiting or hitting-times from the sequence $\mathbf{x}$. The method for extracting samples we use, is similar to the one introduced in Ref.~\onlinecite{rsg2016symbolic}, which is used for estimating the symbolic complexity and particularly, the topological entropy of a process.  After that, we will define the estimators of the entropy rate and entropy production rate, using the fact that the observed samples might be censored.

\subsection{\label{ssec:return}Return time}

\begin{figure}[ht]
\begin{center}
\scalebox{0.4}{\includegraphics{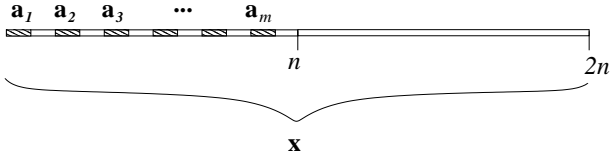}}
\end{center}
     \caption{
 	Selection of sample words for return-time statistics. 
     }
\label{fig:fig01}
\end{figure}

First, we establish the method for obtaining samples of the return-time. Given a sequence $\mathbf{x}$ of size $2n$, take two non-negative integers $\ell$ and $\Delta$ such that $\ell < \Delta  \ll n$. Then define the set $  \mathcal{M}_\ell^{\mathrm{\rho}} =  \{\mathbf{a}_i : \mathbf{a}_i = \mathbf{x}(i\Delta +1,i\Delta+\ell), 0\leq i  < m\}$, where $ m:=\lfloor n/\Delta \rfloor$, of words of length $\ell$ and evenly $\Delta$-spaced along the first half of the trajectory $\mathbf{x}$. In Fig.~\ref{fig:fig01} we show a schematic representation of how the sample words in $\mathcal{M}_\ell^{\mathrm{\rho}} $ are collected from the trajectory $\mathbf{x}$.
%

Next, we define the sample sets of return times $ \mathcal{R}_\ell $ and reversed return times  $ \overline{\mathcal{R}}_\ell $ as follows. First, we associate to each word $\mathbf{a} \in  \mathcal{M}_\ell^{\rho} $ the censored return time, $\rho_{\ell}^{(n)} (\mathbf{a},\mathbf{x})  $, and the reversed return time, $\rho_{\ell}^{(n)} (\overline{\mathbf{a}},\mathbf{x}) $, as follows,
\begin{eqnarray}
\label{eq:censored_rho}
\rho_{\ell}^{(n)} (\mathbf{a},\mathbf{x}) &:=& \inf\{  t > 1\ :\  x_{k+t}^{k+t+\ell-1} = \mathbf{a},\, t\leq n, \mathbf{a} := x_{k}^{k+\ell-1}\},
\nonumber \\
\\
\rho_{\ell}^{(n)} (\overline{\mathbf{a}},\mathbf{x}) &:=& \inf\{  t > 1\ :\  x_{k+t}^{k+t+\ell-1} = \overline{\mathbf{a}}, \, t\leq n,  \overline{\mathbf{a}} := x_{k+\ell-1}^{k}\}.
\nonumber \\
\end{eqnarray}
Observe that $\overline{\mathbf{a}}$ stands for the block $\mathbf{a}$ with its symbols in a reversed order.
Next, $ \mathcal{R}_\ell $ and $ \overline{\mathcal{R}}_\ell $ are defined by
\begin{eqnarray}
\mathcal{R}_\ell &:=&  \{ t \in \mathbb{N} : \rho_{\ell}^{(n)}( \mathbf{a}) = t, \mathbf{a} \in  \mathcal{M}_\ell^{\mathrm{\rho}} \},
\\
\overline{\mathcal{R}}_\ell &:=&   \{ t \in \mathbb{N} : \rho_{\ell}^{(n)}(\overline{\mathbf{a}}) = t, \mathbf{a} \in  \mathcal{M}_\ell^{\mathrm{\rho}} \}.
\end{eqnarray}
It is necessary to stress the fact that the values in the above-defined sample set are not necessarily all of them numerically well-defined (or uncensored). This is because the return-time defined in eq.~(\ref{eq:censored_rho}) is actually censored from above. Notice that  we impose the condition that $\rho_{\ell}^{(n)} $ take a value no larger that $n$. This is imposed by two reasons: on the one hand, we have that the return-time cannot be arbitrarily large due to the finiteness of the trajectory $\mathbf{x}$. And, on the other hand, although it is possible that the return-time for some sample words might be larger than $n$ and still  well-defined, it is not convenient for the statistics. Let us explain this point in more detail. If we take a sample word $\mathbf{a}$ located at the $k$th site, its corresponding return-time can in principle be at most as large as $n+k-\ell$. This happens when the word  $\mathbf{a}$ occurs (by chance) at the  $n+k-\ell$th  site. Since all the sample words in $  \mathcal{M}_\ell^{\rho} $ are located at different sites along $\mathbf{x}$, it is clear that their corresponding return-time values have different upper bounds. Therefore, if we do not impose a homogeneous upper bound, the collection of return-time samples results in inhomogeneous censored data. As we have seen in section~\ref{ssec:estimation_normal}, having a homogeneous bound (homogeneous censored data) is crucial for implementing our estimators. 

\begin{figure}[ht]
\begin{center}
\scalebox{0.34}{\includegraphics{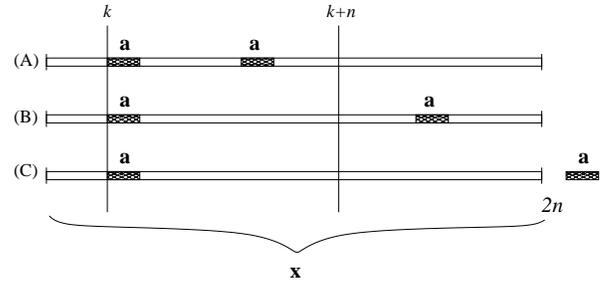}}
\end{center}
     \caption{
Uncensored and censored return-time values. First we suppose that a sample word $\mathbf{a}$ occurs at the $k$th site along a finite trajectory $\mathbf{x}$ of length $2n$. In order to get $\rho_{\ell}^{(n)}(\mathbf{a})$ we should look for the occurrence of $\mathbf{a}$ along $\mathbf{x}$, from the $(k+1)$th symbol to the $(k+n)$th symbol of  $\mathbf{x}$. This section of the trajectory is written $\mathbf{x}(k+1,k+n)$. (A) If $\mathbf{a}$ is found in $\mathbf{x}(k+1,k+n)$, then  $\rho_{\ell}^{(n)}(\mathbf{a})$  is numerically well defined, thus called uncensored. (B) If  $\mathbf{a}$  is found in $\mathbf{x}$ but not in the section $\mathbf{x}(k+1,k+n)$ we consider that  $\rho_{\ell}^{(n)}(\mathbf{a})$ is censored (numerically \emph{undefined} ). (C) Finally, if we do not observe any other occurrence of $\mathbf{a}$ in $\mathbf{x}$ beyond the $(k+1)$th symbol, it is clear that   $\rho_{\ell}^{(n)}(\mathbf{a})$ is numerically undefined, henceforth, censored. 
     }
\label{fig:fig02}
\end{figure}
%

In the following, we will refer to this homogeneous upper bound for return-times as \emph{censoring time} and, whenever convenient it will alternatively be denoted by $T_c$. In Fig.~\ref{fig:fig02} we give an illustrative description of the censoring of the samples.

Once we have the return-time sample set $ \mathcal{R}_\ell $, we introduce the estimator of the entropy rate and the entropy production rate. As we saw in section~\ref{sec:entropy}, if we take a return-time value $t$ from the sample set $ \mathcal{R}_\ell $, then the quantity $\log(t)/\ell$ can be interpreted as a realization of the block entropy rate, $h_{\ell}$  which in the limit when $\ell\to\infty$, obeys the central limit theorem. This fact enables us to implement the following hypothesis: for finite $\ell$, the value $\log(t)/\ell$ is a realization of a normal random variable with (unknown) mean $h_\ell$ and variance $\sigma^2_\ell$. Then, the sample sets 
\begin{eqnarray}
\mathcal{H}_\ell^{\rho} &:=& \{ h = \log(t)/\ell\, : \, t\in  \mathcal{R}_\ell\},
\\
\overline{\mathcal{H}}_\ell^{\rho} &:=& \{ h = \log(t)/\ell\, : \, t\in \overline{ \mathcal{R}}_\ell\},
\end{eqnarray}
can be considered as sets of realizations of normal random variables censored from above by the quantity $h_c:= \log(T_c)/\ell = \log(n)/\ell$ that we call censoring entropy. Then the estimation procedure for the block entropy is essentially the one described in section~\ref{ssec:estimation_normal}. Here we summarize the steps for performing the estimation of $h_\ell$ for return-time statistics. 

\begin{enumerate}

\item Given a finite sample trajectory or a symbolic sequence $\mathbf{x}$ of size $2n$, define the censoring time as the half of the size of the sample trajectory, i.e., $T_c = n$. Fix the number $m$ of sample words or blocks to be collected and the size of the block $\ell$ to be analyzed. Next, define the spacing $\Delta := \lfloor n/m\rfloor$ and the sample set $ \mathcal{M}_\ell^{\mathrm{\rho}} $ of evenly $\Delta$-spaced words that lies along the first half of the trajectory $\mathbf{x}$, i.e.,
\[
\mathcal{M}_\ell^{\mathrm{\rho}} =  \{\mathbf{a}_i : \mathbf{a}_i = \mathbf{x}(i\Delta +1,i\Delta+\ell), 0\leq i < m\}.
\]

\item Define the sets of return-time samples and reversed return-time samples as 
\begin{eqnarray}
\mathcal{R}_\ell &:=&  \{ t \in \mathbb{N} : \rho_{\ell}^{(n)}( \mathbf{a}) = t, \mathbf{a} \in  \mathcal{M}_\ell^{\mathrm{\rho}} \},
\\
\overline{\mathcal{R}}_\ell &:=&   \{ t \in \mathbb{N} : \rho_{\ell}^{(n)}(\overline{\mathbf{a}}) = t, \mathbf{a} \in  \mathcal{M}_\ell^{\mathrm{\rho}} \}.
\end{eqnarray}

\item Using the previous sets of return-time samples define the sets of block entropy and reversed block entropy 
\begin{eqnarray}
\mathcal{H}_\ell^{\rho} &:=& \{ h = \log(t)/\ell\, : \, t\in  \mathcal{R}_\ell\},
\\
\overline{\mathcal{H}}_\ell^{\rho} &:=& \{ h = \log(t)/\ell\, : \, t\in \overline{ \mathcal{R}}_\ell\}.
\end{eqnarray}

\item Next, define the rate uncensored sample values $\hat p := k/m$, where $m$ is the total number of samples in $\mathcal{H}_\ell^{\rho} $ and $k$ is the number of uncensored samples in $\mathcal{H}_\ell^{\rho} $ (henceforth there are $m-k$ censored samples in $\mathcal{H}_\ell^{\rho} $). 

\item Let $1\leq i \leq k$, and denote by $h_i$, each of the uncensored samples in $\mathcal{H}_\ell^{\rho} $. Then its mean and variance are given as follows
\begin{eqnarray}
\bar{h} &:=& \frac{1}{k}\sum_{i=1}^{k} h_i,
\\
s^2 &:=& \frac{1}{k} \sum_{i=1}^{k} (h_i-\bar{h})^2.
\end{eqnarray} 

\item Define the sample functions (see section~\ref{ssec:estimation_normal} and appendix~\ref{ape:censored} for details)
\begin{eqnarray}
\hat \zeta &:=& \frac{\phi(\hat \xi)}{\hat p\, \hat \xi + \phi(\hat \xi)},
\\
\hat \xi &:=& \Phi^{-1}(\hat p),
\end{eqnarray} 
where $\phi(x) = e^{-x^2/2}/\sqrt{2\pi}$ is the probability density function of the standard normal distribution and $\Phi$ its (cumulative) distribution function.

\item Finally, the estimations for the mean of the block entropy and its variance using the return-time estimator are given by
\begin{eqnarray}
\hat h_\ell &=& \bar{h} +\hat \zeta (h_c-\bar{h}),
\\
\hat \sigma^2_\ell &=& s^2 + \hat \zeta (h_c-\bar{h})^2.
\end{eqnarray}
where $h_c$ is the \emph{censoring entropy} and it is defined as
\[
h_c := \log(T_c)/\ell.
\]

\item Repeat steps $4$ -- $7$ for the set $ \overline{ \mathcal{R}}_\ell $ in order to have an estimation of the reversed block entropy rate, which allows to have an estimation of the block entropy production rate just by taking the difference between the reversed block entropy and the 
 block entropy~\cite{gaspard2004time} as follows,
\begin{equation}
\hat{e}_{p} := \hat h_{\ell}^{R}-\hat h_{\ell}.
\end{equation}

\end{enumerate}

\subsection{\label{ssec:waiting}Waiting time}

\begin{figure}[ht]
\begin{center}
\scalebox{0.37}{\includegraphics{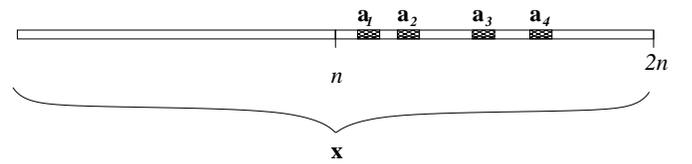}}
\end{center}
     \caption{ Selection of sample words for the waiting time statistics. 
     }
\label{fig:fig03}
\end{figure}

The waiting-time estimator for the block entropy requires two distinct trajectories. In practical situations, we normally have one single trajectory. In order to overcome this problem, we split the original sequence in two equal-sized parts. Since we assume sufficiently rapid mixing, it is possible to regard the second half of the sample to be independent of the first half, provided that the size of the sample is large enough. Thus, one may consider the two parts of the sample as two independent trajectories. After that, we collect $m$ different $\ell$-words at random along one of those trajectories. This collection is denoted by $  \mathcal{M}_\ell^{\omega} $, and will play the role of the set of sample words, in the same way as it was done by set $  \mathcal{M}_\ell^{\rho} $  in section~\ref{ssec:return}. A schematic representation of this sampling procedure is shown in Fig.~\ref{fig:fig03}.
%

The next step consists in defining the censored waiting-time corresponding to each word in the sample $  \mathcal{M}_\ell^{\omega} $. Let $\mathbf{x}$ be the trajectory consisting of $2n$ symbols. Assume that the samples are randomly collected from the segment $\mathbf{x}(n+1,2n-\ell)$. Then we define the censored waiting-time and the censored reversed waiting-time for $\mathbf{a} \in  \mathcal{M}_\ell^{\omega} $ as follows,
\begin{eqnarray}
\omega_{\ell}^{(n)}(\mathbf{a}, \mathbf{x}) &:=& \inf \{ t\geq 1\ :\ x_{t}^{t+\ell-1} = \mathbf{a} \},
\\
\omega_{\ell}^{(n)}(\overline{\mathbf{a}}, \mathbf{x}) &:=& \inf \{ t\geq 1\ :\ x_{t}^{t+\ell-1} =\overline{ \mathbf{a}} \}.
\end{eqnarray}
It is important to notice that the both, the waiting time and the reversed waiting time are bounded from above by $n$, i.e., the sample waiting times are homogeneously censored by $n$. 

The rest of the method follows the lines of the one described in section~\ref{ssec:return}. Here we summarize the main steps:

\begin{enumerate}

\item Given a finite sample trajectory $\mathbf{x}$ of size $2n$, set the censoring time $T_c = n$ equals to the half of the size of the sample trajectory. Fix the number $m$ of sample words to be collected and the size of the block $\ell$. Next, collect $m$ different words at random along the symbolic sequence $\mathbf{x}(n+1:2n)$. We denote by $  \mathcal{M}_\ell^{\omega} $ this collection of $\ell$-words.

\item Define the sets of waiting-time samples and reversed waiting-time samples as 
\begin{eqnarray}
\mathcal{W}_\ell &:=&  \{ t \in \mathbb{N} : \omega_{\ell}^{(n)}(\mathbf{a}, \mathbf{x}) = t, \mathbf{a} \in   \mathcal{M}_\ell^{\omega} \},
\\
\overline{\mathcal{W}}_\ell &:=&   \{ t \in \mathbb{N} :  \omega_{\ell}^{(n)}(\overline{\mathbf{a}}, \mathbf{x})  = t, \mathbf{a} \in  \mathcal{M}_\ell^{\omega}  \}.
\end{eqnarray}

\item From the sets of waiting-time samples define the sets of block entropy and reversed block entropy 
\begin{eqnarray}
\mathcal{H}_\ell^{\omega} &:=& \{ h = \log(t)/\ell\, : \, t\in  \mathcal{W}_\ell\},
\\
\overline{\mathcal{H}}_\ell^{\omega} &:=& \{ h = \log(t)/\ell\, : \, t\in \overline{ \mathcal{W}}_\ell\},
\end{eqnarray}

\item Define the rate of uncensored sample values as $\hat p := k/m$, where $m$ is the total number of samples in $\mathcal{H}_\ell^{\omega} $ and $k$ is the number of uncensored samples also in $\mathcal{H}_\ell^{\omega} $ ( thus, the remaining $m-k$ samples are censored). 

\item Let $1\leq i \leq k$, denote by $h_i$, each of the uncensored samples in $\mathcal{H}_\ell^{\omega} $. Then its mean and variance are given as follows
\begin{eqnarray}
\bar{h} &:=& \frac{1}{k}\sum_{i=1}^{k} h_i,
\\
s^2 &:=& \frac{1}{k} \sum_{i=1}^{k} (h_i-\bar{h})^2.
\end{eqnarray} 

\item Define the sample functions (see section~\ref{ssec:estimation_normal} and appendix~\ref{ape:censored} for details)
\begin{eqnarray}
\hat \zeta &:=& \frac{\phi(\hat \xi)}{\hat p\, \hat \xi + \phi(\hat \xi)},
\\
\hat \xi &:=& \Phi^{-1}(\hat p),
\end{eqnarray} 
where $\phi(x) = e^{-x^2/2}/\sqrt{2\pi}$ is the probability density function of the standard normal distribution and $\Phi$ its (cumulative) distribution function.

\item Finally, the estimations for the mean of the block entropy and its variance using the return-time estimator are given by
\begin{eqnarray}
\hat h_\ell &=& \bar{h} +\hat \zeta (h_c-\bar{h}),
\\
\hat \sigma^2_\ell &=& s^2 + \hat \zeta (h_c-\bar{h})^2,
\end{eqnarray}
again, $h_c$ is the \emph{censoring entropy} defined as above.

\item Repeat steps 4 -- 7 for the set $ \overline{ \mathcal{H}}_\ell^{\omega}  $ in order to have an estimation of the reversed block entropy rate, which allows to have an estimation of the block entropy production rate by taking the difference between the  reversed block entropy and the 
 block entropy~\cite{gaspard2004time} as follows,
\begin{equation}
\hat{e}_{p} := \hat h_{\ell}^{R}-\hat h_{\ell}.
\end{equation}

\end{enumerate}

\subsection{Hitting time}

The hitting-time estimator requires a set of sample words which should be drawn at random from the process that generates the observed trajectory $\mathbf{x}$. Although we do not know the law of the process, we  can still avoid this problem if the set of sample words is obtained by choosing the $\ell$-words at random from another observed trajectory. However, this is the very same method we used for collecting the sample words for the waiting-time estimator. Then,  from the statistical point of view, the hitting-time and waiting-time method can be regarded as the same method. 

\section{\label{sec:tests}Estimations tests}


Now, we will implement the above defined methods for estimating the block entropy and entropy production rates. First of all, we will perform numerical simulations in order to implement a control test statistics which will be compared with the numerical experiments using our methods. 

In section~\ref{sec:estimating_entropy} we established two methods for estimating block entropies by using either, the return-time statistics or the waiting-time statistics. These methods assume that we only have a single ``trajectory'' or, better said, symbolic sequence, obtained by making an observation of real life. Our purpose here  is to test the estimators themselves, and not the sampling methods. The latter means that we will implement the estimators~(\ref{eq:est-h}) and~(\ref{eq:est-sigma}) for both, the return-time and the waiting-time statistics, without referring to the sampling schemes mentioned in section~\ref{sec:estimating_entropy}. This is possible because we have access to an unlimited number of sequences, which are produced numerically with a three-states Markov chain. In this sense we have control of all of the parameters involved  in the estimators, namely, the length of the block $\ell$, the entropy threshold $h_c$ (by which the recurrence-time samples are censored) and the sampling size $|\mathcal{H}_\ell|$.  After that, we will implement the estimation method described in section~\ref{sec:estimating_entropy} using a single sequence obtained from the Markov chain defined below. The latter is a numerical experiment done to emulate an observation of real life where the  accesible sample symbolic sequences are rather limited.

\subsection{\label{ssec:markov-chain}Finite-state Markov chain}

For numerical purposes we consider a Markov chain whose set of states is defined as $\mathcal{A} = \{0,1,2\}$. The corresponding stochastic matrix  $P : \mathcal{A}\times \mathcal{A} \to [0,1]$ is given by,
\begin{equation}\label{eq:stochastic}
P = \left(
  \begin{array}{ccc}
   0 & q & 1-q \\
   1-q & 0 & q \\
   q & 1-q & 0
  \end{array} \right),
\end{equation}
\begin{figure}[t]
\begin{center}
\scalebox{0.37}{\includegraphics{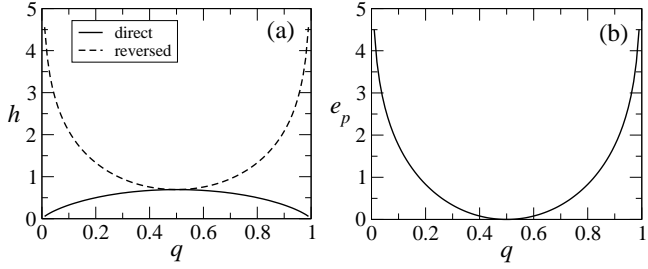}}
\end{center}
     \caption{
     Entropy rate and entropy production rate. (a) We show the behavior of entropy rate $h$  and  reversed entropy rate $h_R$  as a function of the parameter $q$ using the exact formulas given in eqs.~(\ref{eq:h-ex}) and~(\ref{eq:hr-ex}). (b) We display the behavior of entropy production rate as a function of $q$ using the exact formula~(\ref{eq:ep-ex}).
     }
\label{fig:fig04}
\end{figure}
%
where $q$ is a parameter such that $q\in[0,1]$. It is easy to see that this matrix is doubly stochastic and the unique invariant probability vector $\mathbf{\pi} = \mathbf{\pi} P $ is given by $\pi = (\frac{1}{3},\frac{1}{3},\frac{1}{3})$. Moreover, it is easy to compute the entropy rate and the time-reversed entropy rate, indeed, they are given by,
\begin{eqnarray}
h(q)&=& - q\log(q)-(1-q)\log(1-q),
\label{eq:h-ex}
\\
h_R(q)&=&-(1-q)\log(q)-q\log(1-q).
\label{eq:hr-ex}
\end{eqnarray}
Additionally, the corresponding entropy production rate is given by
\begin{equation}
e_p(q) = (2q-1)\log\left( \frac{q}{1-q}\right).
\label{eq:ep-ex}
\end{equation}
The behavior of the entropy rate and entropy production rate can be observed in Figure~\ref{fig:fig04}
We will use this model to generate symbolic sequences in order to test the estimators.

\subsection{Statistical features of estimators for censored data}

The first numerical experiment we perform is intended to show the statistical properties of the estimators without  implementing the sampling schemes introduced above. To this end, we produce a censored sample set of $5\times10^4$ return times obtained from several realizations of the three-state Markov chain. We obtain each of those return times as follows. First we initialize the Markov chain at the stationary state (i.e., we choose the first symbol at random using the stationary vector of the chain) and we make evolve the chain. This procedure generates a sequence which grows in time, say for instance $a_1,a_2,\dots a_t$. The evolution of the Markov chain will be stoped at time $t$ either, until the first $\ell$-word $a_1,a_2,\dots,a_\ell$ appears again, that is if, $ a_{t-\ell+1},a_{t-\ell+2},\dots,a_{t} = a_1,a_2,\dots,a_\ell$ or when the time $t-\ell +1$ exceeds a given bound $T_c$. Then, the corresponding return time (for the $i$th realization) will be either, $\rho_{i} := t-\ell+1$ or an undefined value $\rho_{i}>T_c$.   

%
\begin{figure}[bt]
\begin{center}
\scalebox{0.38}{\includegraphics{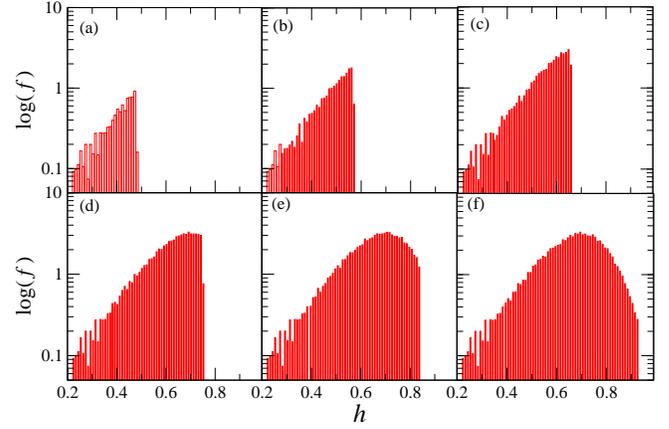}}
\end{center}
     \caption{
     Return-time entropy density for $q=0.60$ and $\ell = 10$.  (a) $h_c = 0.48$,  (b) $h_c = 0.57$,  (c) $h_c = 0.66$,  (d) $h_c = 0.75$, (e) $h_c = 0.84$,  (f) $h_c = 1.02$.
     }
\label{fig:fig05}
\end{figure}
%

Once we have collected the sample set of return times $\{\rho_i\}$ we obtain a set of block entropy rates by means of the equation 
\begin{equation}
h_i  = \frac{\log(\rho_i)}{\ell},
\end{equation}
whenever $\rho_i$ is numerically defined. Of course, we might obtain some numerically undefined sample block entropies $h_i>h_c$ due to the censored return times. 

Analogously, we obtain a sample set of reversed entropy rates. That is, we make evolve the Markov chain and stop its evolution at time $t$ until the first $\ell$-word $a_1,a_2,\dots,a_\ell$ appears reversed in the realization, i.e., $ a_{t-\ell+1},a_{t-\ell+2},\dots,a_{t} = a_\ell,a_{\ell-1},\dots,a_{1}$ or until the time $t-\ell+1$ exceeds the given upper bound $T_c$. The reversed return time for the realization $i$ will be $\rho_{i} = t-\ell +1$ or it is numerically undefined if $t-\ell+1 > T_c$. Then we obtain the sample set  $\{h_i\}$ by means of equation $h_{i} = \log(\rho_{i})/\ell$. 
%
%
\begin{figure}[ht]
\begin{center}
\scalebox{0.35}{\includegraphics{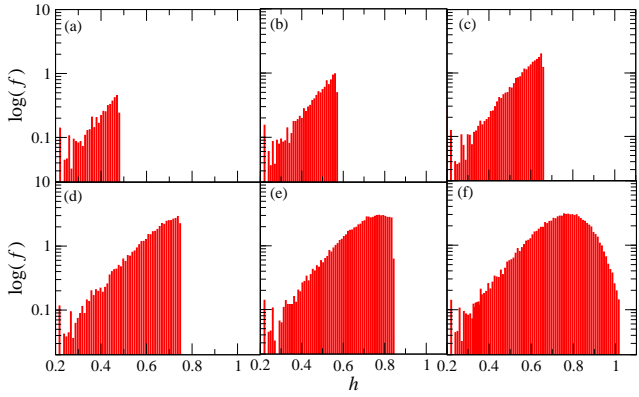}}
\end{center}
     \caption{
          Reversed return-time entropy density for $q=0.60$ and $\ell = 10$.  (a) $h_c = 0.48$,  (b) $h_c = 0.57$,  (c) $h_c = 0.66$,  (d) $h_c = 0.75$, (e) $h_c = 0.84$,  (f) $h_c = 1.02$.
     }
\label{fig:fig06}
\end{figure}
%
%
\begin{figure}[b]
\begin{center}
\scalebox{0.35}{\includegraphics{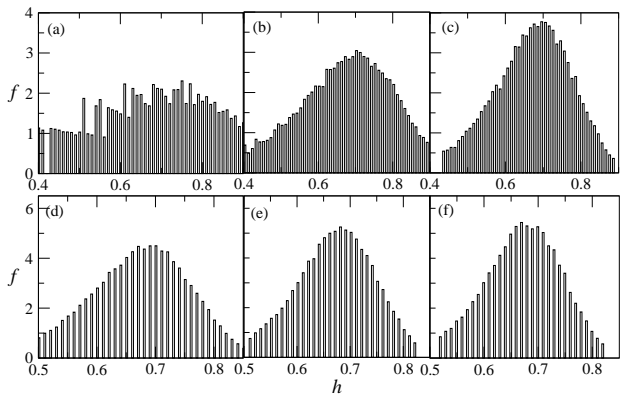}}
\end{center}
     \caption{
         Entropy estimated by means of the return-time statistics for the three-states Markov chain  We show the histograms of the estimated entropy density for $q=0.60$, $h_c = 1.155$ and  (a) $\ell =6$,  (b) $\ell = 9$,  (c) $\ell = 12$,  (d) $\ell = 15$, (e) $\ell = 18$,  (f) $\ell = 19$. We obtained the corresponding histograms using $5\times10^4$ sample words in each case. 
     }
\label{fig:fig07}
\end{figure}
%

Notice that this procedure involves two parameters that can freely vary. These are the block length $\ell$ and  $h_c$ (or equivalently $T_c$), where $h_c$ is an upper bound for the possibly observed block entropy rates, thus, by censoring the corresponding sample set.  

Then, we analyze statistically the sample set of block entropy rates  and reversed block entropy rates  for several values of the free parameters. In Figure~\ref{fig:fig05} we show the histogram of the relative frequencies of the block entropy rate for $\ell = 10$, $q=0.60$ and several values of $h_c$. Correspondingly, in Figure~\ref{fig:fig06} we show the histogram of the relative frequencies of the reversed block entropy rate for $\ell = 10$, $q=0.60$ and several values of $h_c$.

We can appreciate how the density of the block entropy rate is censored while $\ell$ is kept fix. If the value of $h_c$ is small, for most of the samples the return time is numerically undefined because the samples are censored from above. This is seen for instance, in Figure~\ref{fig:fig05}a, in which $h_c$ takes the smallest value for the displayed graphs. In this case, approximately only a $25\%$ of the samples are numerically well-defined resulting in the `partial' histogram displayed in Figure~\ref{fig:fig05}a. In Figure~\ref{fig:fig05}b, the value of $h_c$ is increased causing the histogram to `grow'. In the remaining graphs, from Figure~\ref{fig:fig05}c to Figure~\ref{fig:fig05}d, this tendency is clear, as we increase the value of $h_c$ the number of numerically defined samples grows, thus completing gradually the corresponding histogram. Something similar occurs for the reversed block entropy shown in Figure~\ref{fig:fig06}.

On the other hand, if we keep $h_c$ constant and vary the block length $\ell$, we can appreciate the evolution of the histogram towards a normal-like distribution.  We show this effect in Figure~\ref{fig:fig07} for $q=0.60$ and $h_c = 1.155$ fixed. This is in agreement with the central limit theorem, as we have mentioned in previous sections. In Figure~\ref{fig:fig07} we show the histograms for $\ell = 6, 9, 12, 15, 18$ and $19$ (panels (a)--(f) respectively). We observe that for the lowest value of $\ell$, the histogram is rather irregular, which means that the central limit theorem is still not well manifested for the block entropy rate. We can also observe that increasing the block length, the histogram progressively evolve towards a bell-shaped distribution, which is reminiscent of the normal one. This shows that an estimation using our approach could be more accurate for large values of block lengths due to the central limit theorem. 
%
%
\begin{figure}[t]
\begin{center}
\scalebox{0.22}{\includegraphics{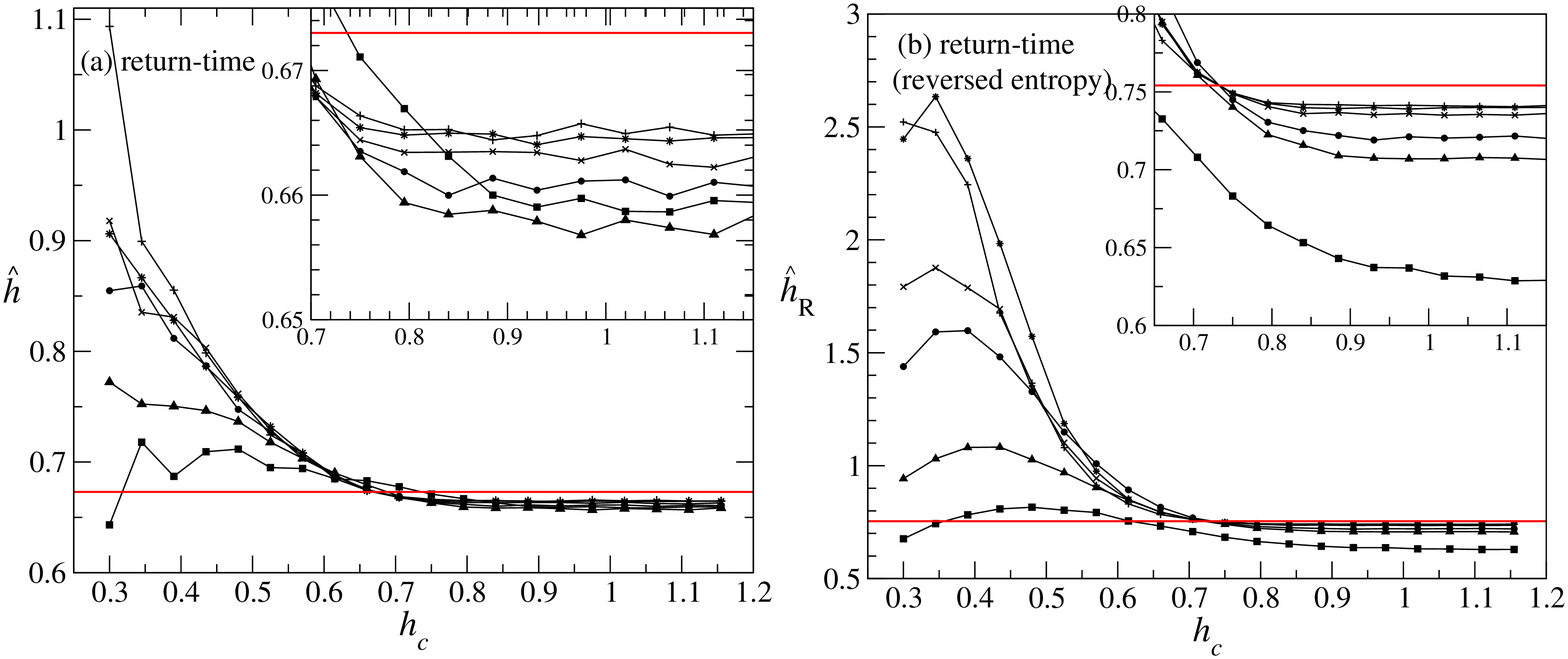}}
\end{center}
     \caption{
     Return-time entropy estimations as a function of $h_c$ for several values of $\ell$. Panel (a): The graphics shows the behavior of $\hat h $ as we increase the entropy threshold $h_c$ for $\ell =6$ (filled squares),  $\ell = 9$ (filled triangles),  $\ell = 12$ (filled circles),   $\ell = 15$ (X's), $\ell = 18$ (stars),  $\ell = 19 $ (plus). Panel (b): it is shown the behavior of the estimated reversed entropy for the same parameter values used in panel (a). 
     }
\label{fig:fig08}
\end{figure}
%

Once we have the sample set of block entropy rates we use the estimation procedure for censored data as described in  Section~\ref{sec:estimating_entropy}. We perform this procedure for the entropy rates and reversed entropy rates obtained from the the return-time and the waiting-time statistics.

In Figure~\ref{fig:fig08} we show the estimation of the block entropy rate and the reversed block entropy rate using the return-time statistics. In Figure~\ref{fig:fig08}a, the displayed curves (solid black lines) show the behavior of the estimation of the block entropy rate as a function of the censoring bound $h_c$ for several values of $\ell$. This figure exhibits two important features  of our estimation technique. Firstly, we notice that the estimation of the entropy rate has large fluctuations for small $h_c$. We can say that the smaller $h_c$, the larger statistical errors are observed, as expected. Secondly, we observe that, the larger $\ell$, the better the estimation. The latter can be inferred from the fact that the curve with the largest value of $\ell$ in Figure~\ref{fig:fig08}a is closest to the exact entropy rate (solid red line). A similar behavior occurs for the reversed block entropy rate estimations shown in Figure~\ref{fig:fig08}b.
\begin{figure}[ht]
\begin{center}
\scalebox{0.235}{\includegraphics{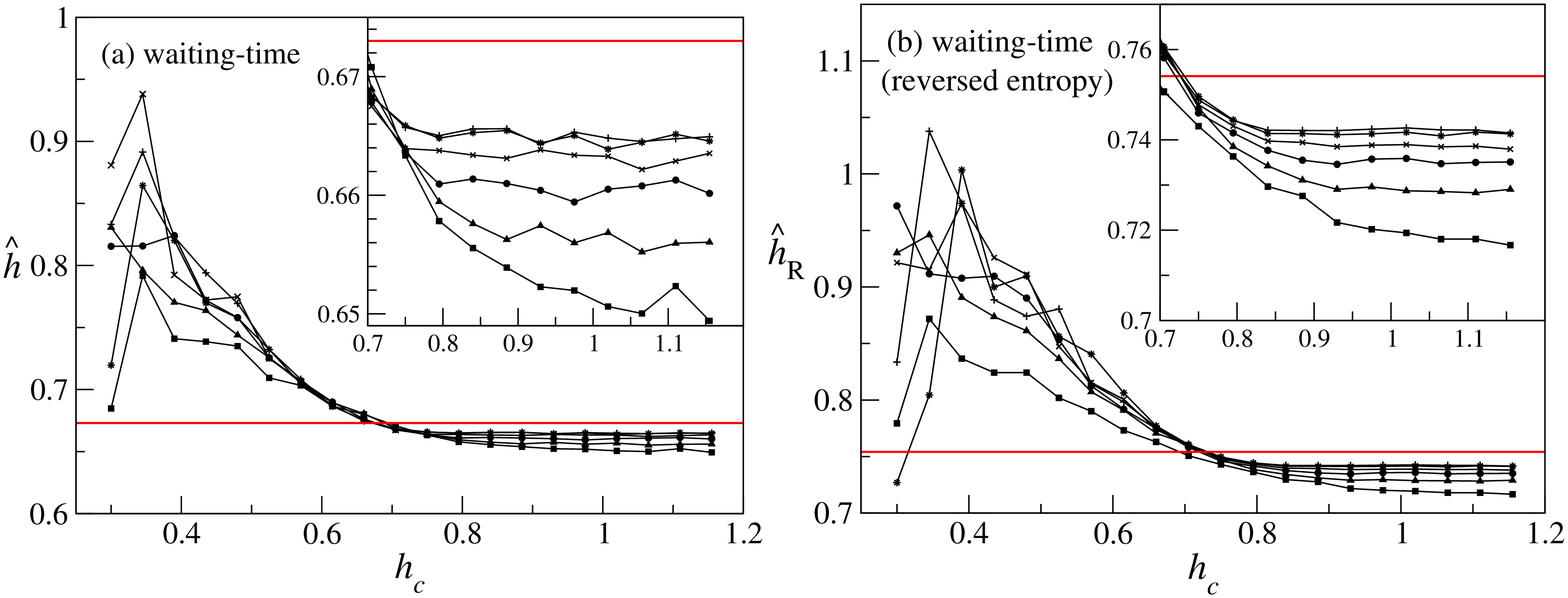}}
\end{center}
     \caption{
     Waiting-time entropy estimations as a function of $h_c$ for several values of $\ell$. Panel (a): The graphics shows the behavior of $\hat h $ as we increase the entropy threshold $h_c$ for $\ell =6$ (filled squares),  $\ell = 9$ (filled triangles),  $\ell = 12$ (filled circles),   $\ell = 15$ (X's), $\ell = 18$ (stars),  $\ell = 19 $ (plus). Panel (b): it is shown the behavior of the estimated reversed entropy for the same parameter values used in panel (a). 
     }
\label{fig:fig09}
\end{figure}
%

For the waiting-time statistics an analogous behavior occurs. In Figure~\ref{fig:fig09} it is shown the curves for the estimations of the block entropy rate, in panel (a), and the reversed block entropy rate, in panel (b). As expected, the estimations for small values of the censoring bound $h_c$ have large fluctuations, which gradually decrease as $h_c$ is  increased. This is clearly observed in Figure~\ref{fig:fig09} because the black solid lines deviate largely  from the exact value (solid red line) for small values of $h_c$. Concerning the value of $\ell$, it is clear that for the largest value of $\ell$, the estimation is closer to the exact entropy rate for $h_c$ large enough (see the insets in Figure~\ref{fig:fig09}).

All these observations allows us to state that, for obtaining the best estimations (as far as possible within the present scheme) we should keep $h_c$ as large as possible. Similarly, in order to assure the central limit to be valid, we should take the block length $\ell$ as large as possible.

Now, we turn our attention to the implementation of the estimations of block entropy rate using the schemes described in Section~\ref{sec:estimating_entropy}. For this purpose, first, we generate a single sequence of $N = 12\times 10^{6}$ symbols by means of the three-states Markov chain. Then, we implement the sampling schemes for the return-time and the waiting-time statistics. In each case, we collect $m=5\times 10^4$ sample words, which correspond to $m=5\times 10^4$ samples of block entropy rates and reversed block entropy rates. These sample sets contain both, numerically defined and undefined samples, the latter ones are due to the censoring. In this case, the censoring bound for entropy rate $h_c$ is determined by
\begin{equation}
\label{eq:h_c-limit}
h_c = \frac{\log(N/2)}{\ell}.
\end{equation}

We should emphasize that in the present case we have control only on a single parameter, which we take as the length  of the block $\ell$. Contrary to the above exposed numerical experiments, in this case $h_c$ is no longer a free parameter; it is actually determined by means of the length of the symbolic sequence $N$ and $\ell$, the length of the block. Consequently, changes in the values of $\ell$ imply changes in the value of $h_c$. The latter is important for two reasons: on the one hand, we have that, in order to assure the validity of the central limit theorem, we should take $\ell$ as large as possible (actually, the true entropy rate is obtained in the limit $\ell \to \infty$). On the other hand, it is desirable to have as much as non-censored samples as possible, i.e., it is convenient for $h_c$ to be as large as possible. However, in practice, we cannot comply with both requirements at once because of expression~(\ref{eq:h_c-limit}): the larger $\ell$, the shorter $h_c$, whenever the length $N$ of the symbolic sequence is kept constant (which commonly occurs in real-world observed data).

\begin{figure}[ht]
\begin{center}
\scalebox{0.37}{\includegraphics{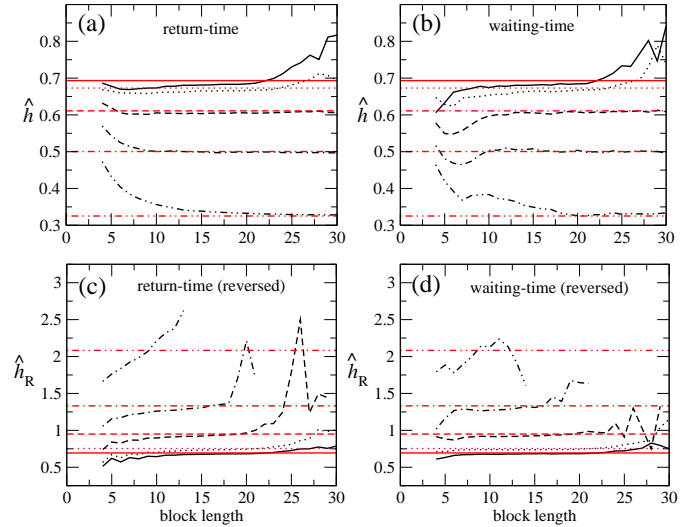}}
\end{center}
     \caption{Estimation of block entropy rate as a function of $\ell$. Black lines stand for the estimated block entropy rate and red lines are the exact entropy rate. We show the curves corresponding to the Markov chain parameter $q=0.50$ (solid lines), $q=0.60$ (dotted lines), $q=0.70$ (dashed lines), $q=0.80$ (dotted--dashed lines)  and  $q=0.90$ (double-dotted--dashed lines). (a) Block entropy rate estimations using the return-time statistics. (b) Same as in (a) using waiting-time statistics. (c) Reversed block entropy rate estimations using the return-time statistics.   (d) Same as in (c) using waiting-time statistics.
     }
\label{fig:fig10}
\end{figure}
%

An important consequence of the latter, that we cannot make $\ell$ as large as we want. Actually, the maximal block length  that it is possible to use for entropy estimations is determined by the accuracy we would like to obtain. This is, for a large $\ell$ we have a short censoring upper bound, implying that only a few samples for block entropy rates are numerically well-defined. This entails a loss of accuracy since, the less numerically well-defined samples, the larger becomes the variance of the estimators. This phenomenon can be observed in Figure~\ref{fig:fig10} for several values of the parameter $q$ of the three-states Markov chain defined in Section~\ref{ssec:markov-chain}.  

In Figure~\ref{fig:fig10}a we show the estimation of block entropy rate as a function of $\ell$ using the return-time statistics. The red lines show the exact value of the entropy rate obtained with eq.~(\ref{eq:h-ex}), while the black lines correspond to the estimations of the block entropy rate using the return-time statistics under the sampling scheme described in Section~\ref{ssec:return}. Figure~\ref{fig:fig10}b shows the same as in Figure~\ref{fig:fig10}a  but using the waiting-time statistics. Figures~\ref{fig:fig10}d  and ~\ref{fig:fig10}d show the corresponding curves for the reversed block entropy rate for the return-time and the waiting-time statistics, respectively. 

%
%
\begin{table}
\caption{\label{tab:return-time-direct}Block entropy estimations using the return-time statistics. }
\begin{ruledtabular}
\begin{tabular}{cc|cccc}
\multicolumn{2}{c}{Parameters }&\multicolumn{4}{c}{Estimations}\\
$q$  &  $\ell^*$  &  $\hat p$  &  $\hat h $ & $\Delta \hat h$  &   $\Delta \hat h/  h $     \\ 
\hline
\hline
$0.5 $&$  22 $&$   0.6066  $&$ 0.692325 $&$  0.000822 $&$  0.001187 $   \\
\hline
$0.6 $&$  23 $&$   0.5488  $&$ 0.669906 $&$  0.003106 $&$  0.004636 $  \\
\hline
$0.7 $&$  25 $&$   0.5718  $&$ 0.607702 $&$  0.003162 $&$  0.005203 $  \\
\hline
$0.80 $&$  30 $&$   0.5880  $&$ 0.496892 $&$  0.003510 $&$ 0.007064  $  \\
\hline
$0.9 $&$  30 $&$   0.9250  $&$ 0.328234 $&$  0.003151 $&$ 0.009600  $   \\
\end{tabular}
\end{ruledtabular}
\end{table}

%
%
\begin{table}
\caption{\label{tab:waiting-time-direct}Block entropy estimations using the waiting-time statistics. }
\begin{ruledtabular}
\begin{tabular}{cc|cccc}
\multicolumn{2}{c}{Parameters }&\multicolumn{4}{c}{Estimations}\\
$q$  &  $\ell^*$  &  $\hat p$  &  $\hat h $ & $\Delta\hat h$  &   $\Delta\hat h/ h $     \\ 
\hline
\hline
$0.5$  &  $22$  &  $ 0.6096$  &  $0.691518$  &  $0.001630$  &  $0.002357   $  \\  
\hline
$0.6$  &  $23 $  &  $0.5460$  &  $0.670477$  &  $0.002535$  &  $0.003781   $  \\
\hline
$0.7$  &  $ 26 $  &  $0.4590$  &  $0.609711$  &  $0.001153$  &  $0.001891   $  \\
\hline
$0.8$  &  $ 30 $  &  $0.5926$  &  $0.496028$  &  $0.004374$  &  $0.008818   $  \\
\hline
$0.9$  &  $ 30 $  &  $0.9242$  &  $0.333756$  &  $0.008673$  &  $0.025986   $  \\
\end{tabular}
\end{ruledtabular}
\end{table}
%
%

Observe that all the curves of the estimated block entropy rate have a common behavior that we anticipated above: there is a special value of $\ell$ for which the estimation seems to be optimal. But, for small and large values of $\ell$ the estimated entropy deviates visibly from the exact value. This phenomenon is produced because, the estimators become better as the $\ell$ increases but also decreases the number of numerically well-defined samples due to the censoring. A criterium for obtaining an optimal $\ell^*$ might not be unique, so here we use a simple one. First of all, once the value of $\ell$ is chosen, the censoring entropy rate is fixed according to eq.~(\ref{eq:h_c-limit}). This bound in turn, determines the number of numerically well-defined samples; the shorter $h_c$, the lower number $k$ of numerically well-defined samples we have. Due to relationship~(\ref{eq:h_c-limit}) we can also say that the larger $\ell$, the lower number $k$ of numerically well-defined samples. A simple way to optimize this interplay between $\ell$ and $k=k(\ell)$, is taking the block length $\ell^*$ for which $k(\ell^*)$  is as close as possible to the half of the sample size $m$.

%
\begin{table}
\caption{\label{tab:return-time-reverse}Time-reversed block entropy estimations using the return-time statistics.  }
\begin{ruledtabular}
\begin{tabular}{cc|cccc}
\multicolumn{2}{c}{Parameters }&\multicolumn{4}{c}{Estimations}\\
$q$  &  $\ell^*$  &  $\hat p$  &  $\hat h_{R} $ & $\Delta \hat h_{R}$  &   $\Delta \hat h_{R}/  h_{R} $     \\ 
\hline
\hline
$0.5$  &  $  22$  &  $ 0.6104$  &  $0.691449$  &  $ 0.001698$  &  $ 0.002456 $ \\ 
\hline
$0.6$  &  $ 21$  &  $ 0.4620$  &  $0.751255 $  &  $0.002850 $  &  $0.003794  $ \\ 
\hline
$0.7$  &  $ 17$  &  $ 0.4504$  &  $0.933973$  &  $ 0.015810$  &  $ 0.016928  $  \\
\hline
$0.8 $  &  $12$  &  $ 0.5586$  &  $1.272741$  &  $ 0.059438$  &  $ 0.046701  $  \\
\hline
$0.9$  &  $ 8$  &  $ 0.4642$  &  $1.981793$  &  $ 0.101070$  &  $ 0.050999  $  \\
\end{tabular}
\end{ruledtabular}
\end{table}
%
%

%
%
%
%
%

%
\begin{table}
\caption{\label{tab:waiting-time-reverse}Time-reversed block entropy estimations using the waiting-time statistics.  }
\begin{ruledtabular}
\begin{tabular}{cc|cccc}
\multicolumn{2}{c}{Parameters }&\multicolumn{4}{c}{Estimations}\\
$q$  &  $\ell^*$  &  $\hat p$  &  $\hat h_{R} $ & $\Delta \hat h_{R}$  &   $\Delta \hat h_{R}/  h_{R} $     \\ 
\hline
\hline
$0.5$  &  $22 $  &  $0.6094$  &  $0.692784$  &  $0.000363$  &  $0.000524   $  \\
\hline
$0.6$  &  $21 $  &  $0.4612$  &  $0.751243$  &  $0.002861$  &  $0.003808   $  \\
\hline
$0.7$  &  $16 $  &  $0.6494$  &  $0.938406$  &  $0.011377$  &  $0.012124   $  \\
\hline
$0.8$  &  $12 $  &  $0.5286$  &  $1.285067$  &  $0.047112$  &  $0.036661   $  \\
\hline
$0.9$  &  $8  $  &  $0.4472$  &  $1.999906$  &  $0.082957$  &  $0.041480   $  \\
\end{tabular}
\end{ruledtabular}
\end{table}
%
%

Using this criterium we compute the optimal block length $\ell^*$, and the corresponding estimated value of entropy rate, for several values of the parameter $q$ of the Markov chain. Tables~\ref{tab:return-time-direct} and~\ref{tab:waiting-time-direct} we show the estimated entropy rate $\hat h$ and the optimal value $\ell^*$ for $q=0.50$, $q=0.60$, $q=0.70$, $q=0.80$,  and $q=0.90$ for the return-time and the waiting-time statistics respectively. We also show a comparison of the estimated block entropy rate with their corresponding exact values.  We can appreciate from these tables that the relative error $\Delta \hat h/h$ (the relative difference between the estimation and the exact value) is lower than $0.06$. Moreover, for $q=0.50$ and $q=0.60$, the relative errors are even less than $1\%$. In Tables~\ref{tab:return-time-reverse} and~\ref{tab:waiting-time-reverse} we show the estimations of reversed entropy rate, and the corresponding optimal $\ell^*$, for the return-time and the waiting-time statistics respectively.

\begin{figure}[ht]
\begin{center}
\scalebox{0.25}{\includegraphics{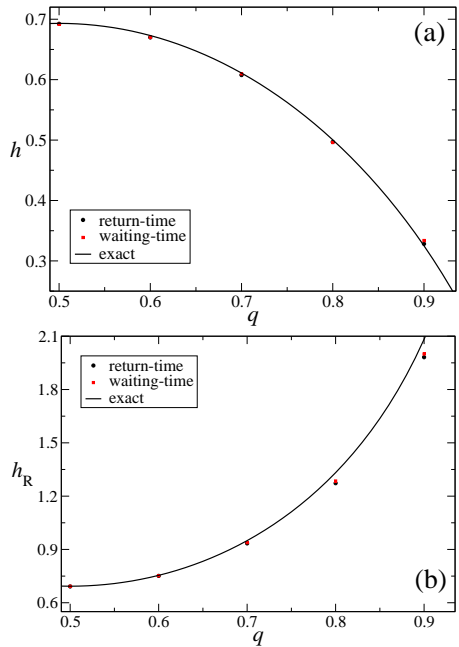}}
\end{center}
     \caption{Estimation of block entropy rate and reversed block entropy rate as a function of $q$. (a) It is shown the block  entropy rate estimated from the return-time statistics (black filled circles) and the waiting-time statistics (red filled squares). We also show the corresponding exact values of the entropy rate (black solid line) of the system for comparing these estimations. In panel (b) the same as in panel (a), but for the reversed entropy rate. 
     }
\label{fig:fig11}
\end{figure}

In Figure~\ref{fig:fig11} we show both the block entropy rate (panel a) and the reversed block entropy rate (panel b) as a function of the parameter $q$. In that figure, the estimation corresponding to the return-time and waiting-time statistics are compared with the exact value. We observe that the return-time and waiting-time statistics have approximately the same accuracy. From Figure~\ref{fig:fig11} we can also see an interesting behavior of the estimation, that is, the larger the entropy rate, the larger the deviation from the exact result. This effect can actually be explained as follows. First we should have in mind that the return-time and waiting-time can be interpreted as measures of the recurrence properties of the system. Specifically, the entropy rate itself can, in some way, be interpreted of as a measure of the recurrence time per unit length of the word (this is a consequence of the fact that the logarithm is a one-to-one function). Thus, it becomes clear that the larger entropy rate, the larger the recurrence times in the system. Since all the samples are censored from above it should be clear that a system having larger recurrence times will have larger errors in the estimations. Therefore we may say that a system with large entropy rate will exhibit large statistical errors in its estimations. Despite of this effect, we observe in  Figure~\ref{fig:fig11}  that the errors in the estimations are sufficiently small for practical applications.

%

Finally we show in Table~\ref{tab:entropy_prodution} the entropy production rate of the system by taking the difference between the block entropy rate and the reversed block entropy rate, for both, the return and the waiting-time statistics. It is important to remark that these recurrence statistics are consistent one to each other, having moderate deviations (statistical errors) when compared with the exact values.

%
%
\begin{table}
\caption{ \label{tab:entropy_prodution} Entropy production estimations from return and waiting time statistics.}
\begin{ruledtabular}
\begin{tabular}{c|cc|cc}
& \multicolumn{2}{c}{Return }&\multicolumn{2}{c}{Waiting}\\
$q$  &     $\hat e_\mathrm{p}$ &  $\Delta e_\mathrm{p}$  & $\hat e_\mathrm{p}$  &  $\Delta e_\mathrm{p}$  \\
\hline
\hline
$ 0.50 $  &  $ -0.000876 $  &   $ 0.000876 $ &  $ 0.001266 $  &   $ 0.001266 $  \\
\hline
$ 0.60 $  &  $ 0.081349 $  &   $ 0.000256 $ &  $ 0.080766 $  &   $ 0.000327 $  \\ 
\hline
$ 0.70 $  &  $ 0.326271 $  &   $ 0.012648 $ &  $ 0.328695 $  &   $ 0.010224 $  \\ 
\hline
$ 0.80 $  &  $ 0.775849 $  &   $ 0.055928 $ &  $ 0.789039 $  &   $ 0.042738 $  \\ 
\hline
$ 0.90 $  &  $ 1.653559 $  &   $ 0.104221 $ &  $ 1.666150 $  &   $ 0.091630 $  \\ 
\end{tabular}
\end{ruledtabular}
\end{table}
%
%

\section{\label{sec:examples}Examples}


In this section we apply our methodology for estimating entropy rate in some well-studied systems for which either, the entropy rate or the (positive) Lyapunov exponents are known. We also include a couple of examples for estimating the entropy production rate for showing the performance of our estimator for the analysis of the time-reversibility (or time-irreversibility) of the process from a finite time-series. 

\subsection{Entropy rate for chaotic maps}

For one-dimensional chaotic maps, a theorem of Hofbauer and Raith~\cite{hofbauer1992hausdorff} allows to compute the entropy rate by means of the Lyapunov exponent and the fractal dimension of the corresponding invariant measure. We use the results reported in Ref.~\onlinecite{calcagnile2010non} for the entropy rate estimated from the Lyapunov exponents as reference values. We test our methodology for three chaotic maps: a Lorenz-like transformation, the logistic map and the Manneville-Pomeau map. 

The first chaotic map we use to exemplify our estimator for entropy rate is a Lorenz-like map $L :[0,1] \to [0,1]$ defined as
\begin{equation}
L(x) :=  \left\{ \begin{array} 
            {r@{\quad \mbox{ if } \quad}l} 
   1-\left( \frac{3-6x}{4}\right)^{3/4}   &   0\leq x < 1/2   \\ 
   \left( \frac{6x-3}{4}\right)^{3/4}  &   1/2 \leq x \leq 1.        \\ 
             \end{array} \right. 
\label{eq:Lx}
\end{equation}
It is clear that this map has a generating partition defined by $\{ [0,1/2),[1/2,1] \}$ allowing a direct symbolization of the time series. 

The second chaotic map we use is the well-known family of logistic maps defined as
\begin{equation}
K_a(x) := ax(1-x).
\end{equation}
We take $a=3.6$, $a=3.8$ and $a=4$ corresponding to the entropy rate (estimated from Lyapunov exponents) reported in Ref.~\onlinecite{calcagnile2010non}. As in the case of the Lorenz-like map, the generating partition is given by $\{ [0,1/2),[1/2,1] \}$ which is the one we use for the symbolization of the time-series.

\begin{figure}[t]
\begin{center}
\scalebox{0.47}{\includegraphics{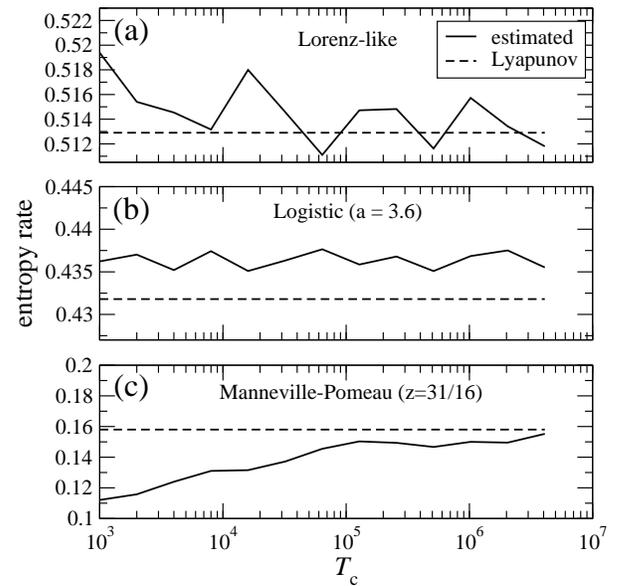}}
\end{center}
     \caption{Estimation of block entropy rate as a function of the censoring time $T_c$ for several chaotic maps. In all these numerical experiments we obtained a symbolic sequence of $8\times 10^6$ symbols long. We obtained a sample set of $2\times 10^4$ words following the waiting time sampling scheme. We fix the censoring time $T_c$ and  compute the corresponding estimations for entropy rate. We repeat the estimation for several values of the censoring time $T_c$. Solid lines represent the waiting-time estimations and dashed lines represent the estimation reported in Ref.~\onlinecite{calcagnile2010non} of the Lyapunov exponent for (a) the Lorenz-like map (b) the logistic map ($a=3.8$) and (c) the Manneville-Pomeau map ($z=31/64$).       }
\label{fig:fig12}
\end{figure}
%

Finally we test our method on the Manneville-Pomeau maps defined as
\begin{equation}
M_z(x):= x+x^z (\mbox{mod } 1), 
\end{equation}
which is a family of chaotic maps exhibiting a dynamics with long range correlations. This family is parametrized by $z\in \mathbb{R}^+$. We concentrate on parameter values within the interval $1<z<2$ for which the map admits a unique absolutely continuous invariant measure~\cite{calcagnile2010non}. Additionally, for such parameter values the dynamics has a power law decay of correlations. We use the parameter values $z_1=3/2$, $z_2= 7/4$, $z_3= 15/8$, $z_4= 31/16$, $z_5= 63/32$, and  $z_6= 127/64$. In this case the natural partition (which is a generating one) is $\{ [0,c_i), [c_i,1]\}$ where $c_i$ is defined as the solution to the equation $M_{z_i} (c_i^-) = 1$.~\cite{calcagnile2010non}

In Ref.~\onlinecite{calcagnile2010non}, the authors estimated the entropy rate by means of the Lyapunov exponent. We take these values as reference for comparison with our results, which are displayed in Table~\ref{tab:entropy_examples}. 

The estimations of entropy rate for these examples are performed as follows. First of all we generate a time-series from the dynamics of each map. Each sequence is symbolized according to the corresponding generating partition. The symbolic sequences obtained from this procedure are $8\times 10^6$ symbols long.  Then, in order to test the performance of our method, we estimate the entropy rate for several values of the censoring time $N$. For a fixed value of censoring time we apply the sampling scheme for the waiting time estimator and we use the criterium given in the preceding section to optimize the block-length $\ell^*$ which the best estimation with respect to the number of samples (recall that $\ell^*$ is the block-length for which the number of samples is closest to the half of the sample words). In this experiments we fix the number of samples to $m=2\times 10^4$ and we estimate the entropy rate for $T_c=2^j\times 10^3$, with $j=1,2,3,\dots 12$. This means that each estimated value of the entropy rate actually does not uses all of the information available in the sample trajectory (except the last one corresponding to $j=12$). This numerical experiment is intended to show the actual performance of our estimators for cases in which the sample trajectories have a ``small'' number of symbols. The estimations of entropy rate as a function of the censoring time $T_c$ are shown in Figure~\ref{fig:fig12}. In that figure we show the estimations for the Lorenz map, the logistic map ($a=3.8$) and the Manneville-Pomeau map ($z=31/64$).

%
%
\begin{table}
\caption{ \label{tab:entropy_examples} Entropy rate estimations for chaotic maps. The estimated entropy rates $h_{\mathrm{Lyap}}$ were obtained from Ref.~\onlinecite{calcagnile2010non} and correspond to the estimation of entropy rate through the Lyapunov exponent.}
\begin{ruledtabular}
\begin{tabular}{ccc|cccc}
 &\multicolumn{2}{c}{ }&\multicolumn{2}{c}{Estimation}\\
 \hline
Map  & Parameter & $\ell^*$  & $h_{\mathrm{Lyap}}$~\cite{calcagnile2010non}& $\hat h$ & $\hat \sigma$ & $\Delta \hat h/h$\\
 \hline
 Lorenz &  & $29$  & $0.5129$  & $0.5134$ & $0.060$  & $10^{-4}$ \\
 \hline
  & $a = 3.6$ &        $53$  & $0.1834$   & $0.1971$ & $0.035$  & $0.074$ \\
 Logistic & $a=3.8 $ & $34$  & $0.4318$   & $0.4375$ & $0.075$  & $0.013$ \\
  & $a=4.0$ &          $22$  & $0.69314$  & $0.6813$ & $0.076$  & $0.017$ \\
\hline
  & $z = 3/2$ &        $25$  & $0.5621$   & $0.5877$ & $0.190$  & $0.046$ \\
  & $z = 7/4$&         $35$  & $0.3597$   & $0.4214$ & $0.226$  & $0.171$ \\
Manneville- &$z = 15/8$& $61$  & $0.2176$  & $0.2384$ & $0.123$  & $0.097$ \\
Pomeau  & $z = 31/16$ &   $91$  & $0.1580$   & $0.1494$ & $0.071$  & $0.054$ \\
  & $z = 63/32$ &         $115$  & $0.1213$   & $0.1141$ & $0.078$  & $0.059$ \\
  & $z = 127/64$ &         $111$  & $0.1164$  & $0.1196$ & $0.075$  & $0.027$ \\
\end{tabular}
\end{ruledtabular}
\end{table}
%
%

The behavior of the estimated entropy rate displayed in Figure~\ref{fig:fig12} shows the performance of our estimations when we vary the censoring time $T_c$. We emphasize that these numerical experiments exhibit certain robustness of our method; for low censoring times (i.e. for $T_c = 1000$ or $T_c=2000$) the deviation of the estimated entropy from the reference value is maintained within the same order of magnitude as the estimated entropy obtained from large censoring times ($T_c=2048000$ or $T_c=4096000$). This behavior means that the proposed estimation methods could, in principle, be implemented in situations in which the sample symbolic sequences are relatively small. It is also important to notice that the way in which the estimated entropy rate behaves is different for every chaotic map. For instance,  the Lorenz-like map seems to converge rapidly with $T_c$, the ``oscillations'' around the reference value might be occasioned by the fluctuations of the estimator itself. Conversely, for the logistic map and the Manneville-Pomeau map the estimated entropy remains above in the former and below in the latter, from the corresponding reference value. This behavior might be the result of several situations that were not be taken into account in the presented study. For example, in the Manneville-Pomeau map we should be aware of the presence of the long-range correlations. These correlations might cause to reach the central limit so slowly in such a way that the proposed estimators might be biased.  This effect can be appreciated in  Table~\ref{tab:entropy_examples} where we summarize the estimated entropy rates for all the chaotic map (and the parameter values) described above using a censoring time $T_c = 1024000$. Here we can see that the half censored samples criterium is meet at large values of the block-length (for $z=63/32$ it is necessary to analyze blocks of length of $115$ symbols). This phenomenon can be explained as follows. Due to the correlations, the symbolic sequences obtained from the symbolization of the chaotic trajectories, might contain large blocks composed, for instance, from a single symbol. Thus any recurrence time associated to these blocks  largely deviates from the typical recurrence time, resulting in a bias in the estimated entropy rate. This clearly deserves a much more detailed study. On the other hand, regarding the estimated results displayed in Table~\ref{tab:entropy_examples}, we should emphasize that  all the reported entropy rates are consistent~\footnote{Notice that Calcagnile \emph{et al}~\cite{calcagnile2010non} report entropy rate values in bits per second because they use logarithm in base 2. Here we use natural logarithm for defining entropy rate and entropy production rate. Thus we should multiply the results of  Calcagnile \emph{et al} by $\ln(2)$ in order to compare them with the ones reported in this work.} with the ones reported in reference~\onlinecite{calcagnile2010non}. We should notice that in this table we display the estimated standard deviation $\hat\sigma$ which is an additional result of our treatment. Since we assume that the estimator has a normal distribution we can estimate how much the estimated entropy rate will deviate from the exact (unknown) value, a deviation that is always present for finite $\ell$ regardless the size of the sample. As we can see in Table~\ref{tab:entropy_examples}, this quantity is consistent with the fact that the reference value is located within the estimated error, taking the error as the estimated standard deviation. Finally in Table~\ref{tab:entropy_examples} we also show the relative error of our estimated entropy rate compared with the reference entropy reported in Ref.~\onlinecite{calcagnile2010non}. These results exhibit the same behavior as we found in analyzing the Markov chain: the larger entropy rate the larger the relative error in our estimations. This phenomenon might also be the result of the fact that for large entropy rate values, the system exhibits a large diversity of blocks of given length. This diversity would imply that, for estimating entropy rate it becomes necessary to have access to a large amount of sample blocks in order to sample accurately all the possible recurrence times and, consequently, to have a sufficient statistics for the block entropy rate estimations.

\subsection{Testing time-irreversibility}

In this section we test the time-irreversibility of a symbolic sequence by using our estimators for entropy rate.  The first example we use is a three-states $n$-step Markov chain which is defined as follows. First, take some $n\in \mathbb{N}$, we  use $n = 6,8,10,12$, and $14$ for our numerical experiments. Define a stochastic process $\{X_t \in \mathcal{A} : t\in \mathbb{N}_0\}$ with state space as $\mathcal{A} =\{0,1,2\}$. The process is initialized as follows. For $0\leq t < n$ we define $X_t$ as an independent random variable with uniform distribution in $\mathcal{A}$. Then, the random variable $X_{n+j}$ is dependent on the random variable $X_j$ in such a way that the conditional probability of $X_{n+j}$ taking the value $a\in \mathcal{A}$  given that $X_j$ took the value $b\in \mathcal{A}$ is defined as
\begin{equation}
\mathbb{P}(X_{n+j} = a | X_j = b) = P_{b,a},
\end{equation}
for all $j\in \mathbb{N}_0$. Here $P$ is a $3\times 3$ square matrix given by the one introduced in eq.~\eqref{eq:stochastic}, i.e.,
\begin{equation}
\label{eq:stochastic-2}
P = \left(
  \begin{array}{ccc}
   0 & q & 1-q \\
   1-q & 0 & q \\
   q & 1-q & 0
  \end{array} \right).
\end{equation}
The case $n=1$ reduces to the 1-step three-states Markov chain introduced in section~\ref{sec:tests} for testing the entropy estimators. It is clear that, by construction, the irreversibility of the process will be manifested when analyzing block lengths larger than $n$, no matter which method is used for estimations.  This is because each realization of the $n$-step Markov chain can be interpreted  as a $n$-independent realization of the $1$-step Markov chain.  In other words,  for $0\leq k < n$, we can interpret the processes  $\{X_{nj+k} : j\in \mathbb{N}\}$ as a collection of $n$ realizations of a $1$-step Markov chain given by the stochastic matrix $P$. So, in order to see the effects of the irreversibility it is necessary to analyze blocks with size beyond $n$. 
%
\begin{figure}[ht]
\begin{center}
\scalebox{0.35}{\includegraphics{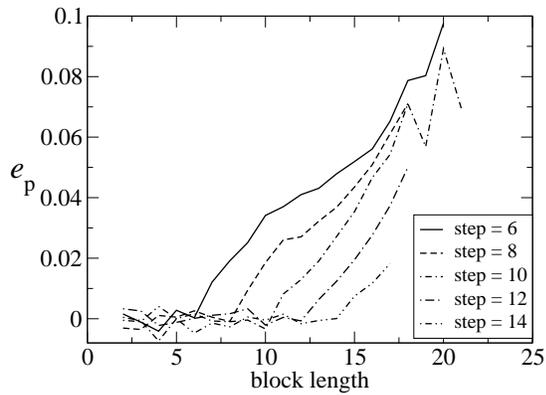}}
\end{center}
     \caption{Estimation of block entropy production rate as a function of $\ell$ for a $n$-step Markov chain.  For these estimations we used a sequence of $8\times 10^6$ symbols long. We collected a sample set of $2\times 10^4$ words of size $\ell$. We vary $\ell$ from $\ell = 2$ up to the value of $\ell$ such that the number of censored samples exceed $80\%$. We make these estimations for the $n$-step Markov chain for $n=6,8,10, 12$ and $14$.     }
\label{fig:fig13}
\end{figure}
%

For the numerical experiments we perform a realization of the $n$-step Markov chain obtaining a symbolic sequence of $8\times 10^6$ symbols long for $n=6,8,10, 12$, and $14$ and we fix $q=0.60$. Then we use this sequence to obtain an estimation of entropy production rate. We perform such estimations by implementing the sampling scheme for the waiting-time statistics with a censoring time $T_c = 4\times 10^6$. The total number of collected samples was $2\times 10^4$. The entropy production rate was obtained as the difference between the reversed entropy rate and the entropy rate, as a function of the block-length $\ell$ starting at $\ell = 2$ and increasing this value successively.  The algorithm is stopped when the block-length $\ell$ is such that the number of censored samples exceeds $80\%$. This  estimation of the entropy production rate could be used as an irreversibility index which can be appreciated in Figure~\ref{fig:fig13}.  We can see that the estimated entropy production rate is consistent with the fact that no entropy production is present while analyzing block-lengths smaller that the order of the Markov chain. Notice that, as shown in Figure~\ref{fig:fig13}, the estimated entropy production rate is, in average, zero for block-lengths below $n$ and that above such value the estimated entropy production rate gives a positive value.

%
\begin{figure}[t]
\begin{center}
\scalebox{0.4}{\includegraphics{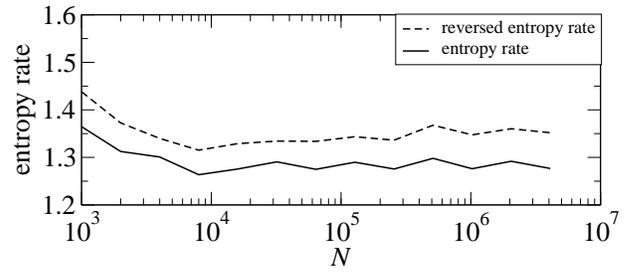}}
\end{center}
     \caption{Estimation of block entropy rate as a function of $\ell$.      }
\label{fig:fig14}
\end{figure}
%

Another interesting application is the analysis of the irreversibility of DNA. In Ref.~\onlinecite{Provata2014} Provata \emph{et al} explored the irreversibility of some chromosomes of human DNA. Particularly they use the empiric measure to estimate the entropy rate and the reversed entropy rate to give an estimation of the degree of irreversibility of the chromosomes 10, 14 20 and 22. The estimation of the entropy rate they report was done under the assumption that the genomic sequences are produced by a $4$-step Markov chain.~\cite{Provata2014} They report the values for entropy rate $\hat h^* = 1.339$ and $\hat h_R^* = 1.416$ (see Table VI in Ref.~\onlinecite{Provata2014}).  

Here we obtain from the GenBank~\cite{benson1997genbank} the chromosome 20 of the \textit{Homo sapiens} and we use this sequence to estimate entropy rate and the reversed entropy rate using the sampling scheme corresponding to waiting time statistics given in section~\ref{sec:estimating_entropy}. Although chromosome 20 has $63\times 10^6$ base pairs (symbols) we only use a segment of $8\times 10^6$ base pairs long to achieve our estimators. We actually estimate the entropy rates for several values of the censoring time $T_c$. 

These values are $2^j\times 10^3$, for $0\leq j \leq 12$. The purpose of these estimations is to see how much the estimations deviate when using a ``small sample'' in the sense of the size of the observed trajectory. We fix the number of sample words to $m = 2\times 10^4$.  We also use the criterium given in the preceding section in order to optimize the block-length $\ell^*$ which gives the best estimation with respect to the number of censored samples.
 
In Figure~\ref{fig:fig14} we display the behavior of $\hat h$ and $\hat h_R $ as a function of the censoring time $T_c$. We can appreciate that the entropy rate and the reversed entropy rate has a difference which is, the average, the same for all the values of the censoring times implemented. This fact suggest  that the DNA sequence might be spatially irreversible in the sense of the positivity of the entropy production rate. These results are compared with the values for the entropy rate and the reversed entropy rate reported in Ref.~\onlinecite{Provata2014}. For $T_c = 4096000$ we obtain $\hat h = 1.277$ and $\hat h_R =1.352$ which are, up to some extend, consistent with the values obtained in  Ref.~\onlinecite{Provata2014} since we obtain an entropy production rate $\hat e_p = 0.075$ while the one reported in Ref.~\onlinecite{Provata2014} is $\hat e_p^* = 0.077$.

\bigskip

\section{Conclusions}
\label{sec:conclusions}

The entropy rate is the limit of the block entropy rate when the block length goes to infinity. Attaining the limit of infinite block length is impossible in practice. Moreover, estimating block entropy rate from empiric measures would require a large amount of data even for moderately large  block lengths.  Furthermore, in this work we have shown that the finiteness of the observed trajectories may lead to errors that can be associated to censored samples of the entropy rate. 

We have studied estimators for the entropy rate defined from the recurrence-time statistics; specifically the return-time and waiting-time estimators. Taking into account the problem of the finiteness of the observed trajectory, we have used the theory of censored samples from statistics in order to obtain improved estimators for the entropy rate. From this point of view, we established a couple of sampling schemes for the return-times and waiting-times in order to implement the corresponding maximum likelihood estimators for the censored normal distribution. The latter is justified by assuming that the entropy rate estimator comply with the central limit theorem. These results show that there is some compromise between the length of the words used for the estimation and the size of the sample it self. This must be considered in order to obtain the optimal estimation for a given sample. The protocols we define are, in some sense, a new technique since we take advantage of the approach of the recurrence-time statistics for estimating the entropy rate (and entropy production rate) combined with the existing tools for censored data statistics.

In view of the examples analyzed for the case of one-dimensional maps with chaotic dynamics we can say that the proposed methodology is at  least as accurate as the existing tools for estimating entropy rates. Moreover, we have also shown through two simple examples (an $n$-step Markov chain and a real DNA sequence) that our method could be a useful tool for detecting reversibility or irreversibility in a time series.

We would like to highlight the importance of this approach, since it might be applicable for more general systems than Markov chains. The main assumptions we have made are applicable to more general dynamical systems and stochastic processes satisfying the appropriate mixing properties. Nevertheless, for the specific case of entropy production rate, if one uses directly the protocol here defined, one would obtain estimators of certain indexes of irreversibility instead of the true entropy production rate. Finally, we have defined applicable protocols for time series taken from real data, and thus, important for practical purposes, as we have shown for the case of DNA. 

\bigskip

\begin{acknowledgments}
This work was supported by CONACYT through grants A1-S-15528 and FORDECYT-PRONACES/1327701/2020.
RSG thanks the Instituto de F\'isica–UASLP as well as the Instituto Potosino de Investigaci\'on Cient\'ifica y Tecnol\'ogica (IPICyT) for their warm hospitality. 
\end{acknowledgments}

\appendix

\section{Maximum Likelihood estimators for normal censored samples.}
\label{ape:censored}

In this appendix we derive formulas~(\ref{eq:hat_h}) and~(\ref{eq:hat_sigma2}) for estimating the mean and variance (respectively) of the entropy rate assuming a normal distribution. These formulas essentially correspond to the maximum likelihood estimations of the mean and the variance of a normal distribution with samples censored from above. Although the derivation of these estimators are found in Ref.~\onlinecite{cohen1991truncated}, we include the following calculations for the sake of completeness of the present work.

Let $\Theta$ be a random variable normally distributed with mean $h$ and variance $\sigma^2$. Let $\mathcal{H}:= \{h_i : 1\leq i \leq m\}$ be a sample of independent realizations of $\Theta $ censored from above, i.e., a given sample $h_i $ is either, uncesored in the sense that it has a specific numerical value,  or censored, that is, $h_i$ has a value larger than the censoring threshold $h_\mathrm{c}$. We order the sample set in such a way that the first $k$ ($k\leq m$) samples are uncensored ones, i.e. $h_i$ is uncensored for $1\leq i \leq k$ and censored if $k+1\leq i \leq m$. Let us also denote by $\hat p := k/m$ as the fraction of uncensored samples in the whole sample set $\mathcal{H}$. Notice that $\hat p$ is an estimation of the probability of the sample to be uncensored. Since $\Theta $ is assumed to be normal we have that 
\begin{equation}
\label{eq:def:p}
p := \Phi\left( \frac{h_c-h}{\sigma }\right),
\end{equation}
where $\Phi$ is the distribution function of a standard normal random variable, i.e.,
\begin{equation}
\Phi (x) := \frac{1}{\sqrt{2\pi}}\int_{-\infty}^x  e^{-y^2/2}dy.
\end{equation}

Next, the likelihood function, which can be interpreted as the probability of the occurrence of the collected samples, is given by   
\begin{equation}
L(h,\sigma^2; \mathcal{H}) := \left( \prod_{i=1}^k \left( \frac{\Delta h}{\sigma} \phi\left(\frac{h_i-h}{\sigma} \right) \right) \right) \left( 1-p \right)^{m-k},
\end{equation}
where $\phi$ is the probability density function of the standard normal distribution, $\phi(x) := e^{-x^2/2}/\sqrt{2\pi}$. It is not hard to see that the logarithm of the likelihood function (some times also called \emph{log-likelihood function})  can be written as
\begin{eqnarray}
\log L &:=&\left( - \sum_{i=1}^k  \frac{(h_i-h)^2}{2\sigma^2} \right)  + k \log\left(\frac{\Delta h}{\sqrt{2\pi }} \right) 
\\
\nonumber
&-& \frac{k}{2}\log(\sigma^2)  + (m-k)\log\left( 1- \Phi\left( \frac{h_c-h}{\sigma }\right) \right).
\end{eqnarray}

Now, in order to obtain the maximum likelihood estimations we need to maximize the log-likelihood function with respect to the parameters $h$ and $\sigma^2$. After some calculations it is possible to see that the first derivatives of $\log L$ with respect to $h$ and $\sigma^2$ are given by
\begin{eqnarray}
\frac{\partial \log L }{\partial h} &=& \sum_{i=1}^k  \frac{h_i-h}{\sigma^2} + \left( \frac{ m-k}{\sigma}\right)\frac{ \phi\left( \frac{h_c-h}{\sigma}\right) }{ 1- \Phi\left( \frac{h_c-h}{\sigma }\right) },
\\
\frac{\partial \log L }{\partial \sigma^2} &=&  \sum_{i=1}^k  \frac{(h_i-h)^2}{2\sigma^4} -\frac{k}{2\sigma^2} 
\nonumber
\\
&+&
(m-k)\left(\frac{h_c-h}{2\sigma^3 }\right)\frac{ \phi\left( \frac{h_c-h}{\sigma}\right) }{ 1- \Phi\left( \frac{h_c-h}{\sigma }\right) }.
\end{eqnarray}

To maximize $\log L$ we have to equate to zero the above partial derivatives. The solutions will correspond to the maximum likelihood estimations for the mean and variance of the distribution. Then we can write,
\begin{eqnarray}
\label{eq:int-est-1}
 &&\sum_{i=1}^k  \frac{h_i-h}{\sigma^2} + \left( \frac{ m-k}{\sigma}\right)\frac{ \phi\left( \frac{h_c-h}{\sigma}\right) }{ 1- \Phi\left( \frac{h_c-h}{\sigma }\right) }=0,
\\
\label{eq:int-est-2}
 && \sum_{i=1}^k \frac{(h_i-h)^2}{2\sigma^4} -\frac{k}{2\sigma^2} +
(m-k)\left(\frac{h_c-h}{2\sigma^3 }\right)\frac{ \phi\left( \frac{h_c-h}{\sigma}\right) }{ 1- \Phi\left( \frac{h_c-h}{\sigma }\right) } = 0.
\nonumber
\\
\end{eqnarray}

For further calculations it is important to have a short-hand notation, then we define the following quantities. First we denote by $\bar h$ and $s^2$ the sample mean and variance, respectively, as
\begin{eqnarray}
\bar{h} &:=& \frac{1}{k}\sum_{i=1}^{k} h_i,
\\
s^2 &:=& \frac{1}{k} \sum_{i=1}^{k} (h_i-\bar{h})^2.
\end{eqnarray}
Next we denote by $\xi$ the following, 
\begin{eqnarray}
\label{eq:def:xi}
\xi &:=& \frac{h_c-h}{\sigma}.
\end{eqnarray}

In terms of the above quantities it is possible to see that equations~(\ref{eq:int-est-1}) and~(\ref{eq:int-est-2}) can be rewritten as
\begin{eqnarray}
\label{eq:estimation-int-1}
 \frac{\bar{h} - h}{\sigma} + \left( \frac{1-p}{p }\right)\frac{ \phi\left( \xi \right) }{ 1- \Phi\left(\xi \right) }=0, \qquad 
\\
\label{eq:estimation-int-2}
\frac{s^2 + (\bar{h}-h)^2}{2\sigma^2} -\frac{1}{2} +
\left(\frac{(1-p)\xi}{2p }\right)\frac{ \phi\left( \xi\right) }{ 1- \Phi\left( \xi \right) }= 0.\qquad
\end{eqnarray}

Equations~(\ref{eq:estimation-int-1}) and~(\ref{eq:estimation-int-2}) can be further simplified as follows. First let us denote by $\Omega$ the combination
\begin{equation}
\Omega := \left( \frac{1-p}{p }\right)\frac{ \phi\left( \xi \right) }{ 1- \Phi\left(\xi \right) }.
\end{equation}
Then equations~(\ref{eq:estimation-int-1}) and~(\ref{eq:estimation-int-2}) can be rewritten as
\begin{eqnarray}
\label{eq:estimation-int-11}
h-\bar{h} =\sigma \Omega,
\\
\label{eq:estimation-int-22}
s^2 + (\bar{h}-h)^2 = \sigma^2\left[ 1- \xi \Omega \right].
\end{eqnarray}
Now we can use equation~(\ref{eq:estimation-int-1})  into equation~(\ref{eq:estimation-int-2}) to eliminate the dependence on $h-\bar{h}$. This results in
\begin{equation}
\label{eq:s2-int-1}
s^2 = \sigma^2\left[1- \xi \Omega - \Omega^2\right],
\end{equation}
or, equivalently, as 
\begin{equation}
\label{eq:sigma2-int-1}
\sigma^2 = s^2 +  \sigma^2\left[\xi \Omega + \Omega^2\right],
\end{equation}
On the other hand, recalling the definition of  $\xi$, we can write $h = h_c-\sigma \xi$. Using this identity into equation~(\ref{eq:estimation-int-1}) we obtain
\begin{equation}
\label{eq:sigma-int-1}
\sigma = \frac{h_c - \bar{h} }{\left( \xi + \Omega \right)} . 
\end{equation}
Using the last identity into equation~(\ref{eq:sigma2-int-1}) we obtain 
\begin{equation}
\label{eq:sigma2-int-2}
\sigma^2 = s^2 +\frac{ \xi \Omega + \Omega^2}{\left( \xi + \Omega \right)^2} \left(h_c- \bar{h}\right)^2
=  s^2 +\frac{ \Omega }{ \xi + \Omega } \left(h_c- \bar{h}\right)^2.
\end{equation}
The identity~(\ref{eq:sigma-int-1}) can also be used into equation~(\ref{eq:estimation-int-11}) which results in
\begin{equation}
\label{eq:h-int-2}
h =\bar{h} + \frac{\Omega }{\left( \xi + \Omega \right)} \left( h_c - \bar{h} \right).
\end{equation}
Next, we denote by $\zeta$ the combination $\Omega/(\xi+\Omega)$, a quantity which appears in the expression for $\sigma^2$ and $h$. Recalling that $p = \Phi(\xi)$, which is a consequence of expressions~(\ref{eq:def:p}) and~(\ref{eq:def:xi}), we can see that $\Omega = \phi(\xi)/p$. Then, some algebraic manipulations show that
\begin{equation}
\zeta := \frac{ \Omega }{ \xi + \Omega} = \frac{ \phi(\xi)/p }{ \xi + \phi(\xi)/p},
\end{equation}
or, equivalently,
\begin{equation}
\label{eq:zeta-final}
\zeta = \frac{ \phi(\xi) }{ p\xi + \phi(\xi)}.
\end{equation}
In terms of $\zeta$ we have that 
\begin{eqnarray}
\label{eq:h-final}
h = \bar{h} + \zeta \left( h_c - \bar{h} \right),
\\
\label{eq:sigma2-final}
\sigma^2 = s^2 + \zeta \left( h_c - \bar{h} \right)^2.
\end{eqnarray}
Equations~(\ref{eq:zeta-final}), (\ref{eq:h-final}), and~(\ref{eq:sigma2-final}) are the expressions anticipated in section~\ref{ssec:estimation_normal}.

\section*{Data Availability}
The data that supports the findings of this study are available within the article.

\section*{References}
\nocite{*}
\bibliography{Refs_EntropyProduction.bib}

\end{document}